\RequirePackage{fix-cm}
\documentclass[twocolumn]{svjour3}          
\smartqed  
\usepackage{graphicx}
\usepackage{graphics, cite} 
\usepackage{epsfig} 
\usepackage{mathptmx} 
\usepackage{times} 
\usepackage{amsmath} 
\usepackage{amssymb}  
\usepackage{color}
\begin{document}

\title{Chaotic Dynamics of the Chaplygin sleigh with a passive internal rotor \thanks{NSF CMMI grant}
}


\author{Vitaliy Fedonyuk       \and
       Phanindra Tallapragada 
}


\institute{Vitaliy Fedonyuk\at
              \email{vfedony@g.clemson.edu}           
           \and
           Phanindra Tallapragada \at
               \email{ptallap.clemson.edu}  
}

\date{Received: date / Accepted: date}

\maketitle

\begin{abstract}
The Chaplygin sleigh is a classic example of a nonholonomically constrained mechanical system.  The sleigh's motion always converges to a straight line whose slope is entirely determined by the initial configuration and velocity of the sleigh. We consider the motion of a modified Chaplygin sleigh that contains a passive internal rotor. We show that the presence of even a rotor with small inertia modifies the motion of the sleigh  dramatically. A generic trajectory of the sleigh in a reduced velocity space exhibits two distinct  transient phases before converging to a chaotic attractor. We demonstrate this through numerics.  In recent work the dynamics of the Chaplygin sleigh have also been shown to be similar to that of a fish like body in an inviscid fluid. The influence of a passive degree of freedom on the motion of the Chaplygin sleigh points to several possible applications in controlling the motion of the nonholonomically constrained terrestrial and aquatic robots.

\keywords{nonholonomic systems \and passive internal degrees of freedom \and Chaplygin sleigh \and chaotic attractor}
\end{abstract}

\section{Introduction}
Internal degrees of freedom occur naturally in many biological systems. A classic example is of a falling cat that uses internal degrees of freedom between the joints of its spine to consistently land on its feet, \cite{kane_fallingcat_1969, montgomery_1993}. Internal degrees of freedom in the form of  internal momentum wheels also find use in the attitude control of satellites, \cite{tsiotras_2001}. In both of these examples, however, the internal degrees are actuated and can be useful for closed loop control.  In contrast passive degrees of freedom cannot be actuated but can add a passive control layer embodied within the system. Such a phenomenon is observed in many flying insects that use passive twisting of the wing along the leading edge to generate lift during hovering flight, \cite{Dickinson_1999, Whitney_JFM_2010}. 

In the context of application to robots, it is important to extend the concept of passive degrees of freedom to mechanical systems with nonholonomic constraints, which are ubiquitous in many wheeled, legged and snake like robots, \cite{murray_sastry_1993, tpl_2009}. When a mechanical system is nonholonomically constrained, the motion of internal degrees of freedom do change the linear and angular momentum of the system. Nonholonomic systems with even unactuated or passive internal degrees of freedom can behave very differently due to the  exchange of momentum between the different degrees of freedom. In this paper we study the motion of a Chaplygin sleigh, a well known nonholonomic system \cite{Chaplygin_1911,Neimark1972, borisovmamaev02, bloch03}, with the addition of a passive unbalanced internal rotor. Beginning with any initial velocity and angular velocity, the dynamics of the Chaplygin sleigh are such that the angular velocity converges to zero and the sleigh asymptotically attains a constant velocity, while preserving the kinetic energy. The sleigh's trajectory in the plane would converge to a straight line. 

The addition of a `small' internal passive rotor changes the dynamics in two ways, the velocities of the sleigh and the internal rotor converge to a chaotic attractor and a trajectory of the sleigh in the plane remains bounded.  This significant effect that a passive internal degree of freedom can have on a nonholonomic system  inspires questions of passive mechanical control of robotic systems. The addition of dissipative and stiff elements can produce limit cycles such as those observed in \cite{ft_nd_2018} which could passively direct the dynamics of the robot to a desired invariant state. A further significant feature of the dynamics of systems with internal degrees of freedom is that the linear and or the angular momentum of the system is unaffected by the motion of an internal degree of freedom.  

An additional motivation for our investigation into the role of a passive degree of freedom is due to the close analogy between the dynamics of a Chaplygin sleigh and a fish-like body moving in an inviscid fluid. It has been shown recently that the Kutta Joukowski condition that determines vortex shedding past a sharp tip of a body in an inviscid fluid is equivalent to an affine nonholonomic constraint on the motion of the body, \cite{tallapragada_ACC2015, tallapragada_jcnd2016}. It has also been shown that both a Chaplygin sleigh and a fish like body could be propelled and steered by the periodic motion of an internal rotor, \cite{tallapragadakelly13acc, tallapragada_epjst2015, pt_tmech_2016, pt_dscc2016}. More interestingly small passive tail like segments have been shown to dramatically improve the  rapid turning maneuvers of such a swimming robot, \cite{pt_tmech_2018}. Understanding the influence of a passive internal rotor on the motion of the Chaplygin sleigh can aid in the understanding of the role played by passive segments in the locomotion of a fish like body in water. 

This paper is organized as follows: in section \ref{sec:eom} we derive the equations of motion for the Chaplygin sleigh with an internal degree of freedom. In section \ref{sec:simulation} we present some results of the simulation of the equations of motion to highlight the differences in the dynamics due to the passive internal rotor.  In section \ref{sec:transient} we  analyze the transient phase of the motion of the sleigh using a regular perturbation expansion of the equations of motion.  In section \ref{sec:chaos} we explore the convergence of trajectories to a chaotic attractor through numerical simulations.

\section{Equations of Motion}\label{sec:eom}
\begin{figure}[!h]
	\begin{center}
			\includegraphics[width=0.8\hsize]{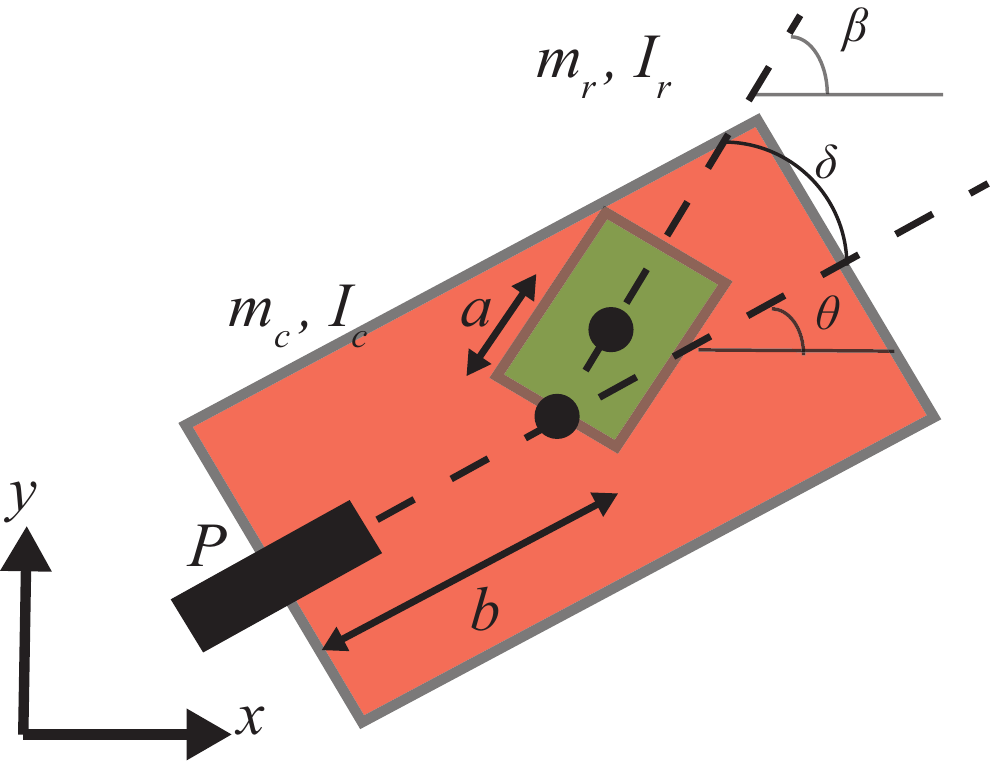}
	\end{center}
	\caption{The Chaplygin sleigh consists of a sleigh of mass $m_c$  with a rear wheel or a sharp runner at a distance of $b$ from the center of mass. The runner makes contact with the ground at point $P$. An internal rotor of mass $m_r$ is attached to the center of the sleigh. The center of mass of the rotor is at a distance of $a$ from the center of the sleigh.}\label{fig:ref_frames}
\end{figure}

A schematic of the Chaplygin sleigh is shown in Fig. \ref{fig:ref_frames}. The sleigh has a runner or a slender wheel at the rear that contacts the ground at the point $P$. The runner is assumed to able to slide smoothly in its longitudinal direction but not in a transverse direction. The mass of the sleigh is denoted by $m_c$ and the moment of inertia about its center by $I_c$. An internal rotor of mass $m_r$ and moment of inertial $I_r$ are hinged to the center of the sleigh, such that the rotor can rotate freely without any resistance. The configuration of the Chaplygin sleigh is parameterized by the location its the center of mass, $(x, y)$ and its orientation $\theta$, with respect to an inertial frame of reference. The configuration of the internal rotor an be parameterized by the angle $\beta \in S^1$.   The configuration space of the system is $Q = SE2 \times S^1$. The tuple $(x,y,\theta, \beta)$ will be  represented by $q = (q_1,q_2,q_3,q_4)$ for convenience. The body axes attached to point $C$ are denoted by $X_b- Y_b$.

The Lagrangian for the system can be written as
\begin{equation}\label{eq:Lagrangian}
\mathcal{L} = \frac{1}{2}m_{c}(\dot{x}^2+\dot{y}^2)+\frac{1}{2}m_{r}(\dot{x}_r^2+\dot{y}_r^2) \notag 
+\frac{1}{2}I_{c}\dot{\theta}^2+\frac{1}{2}I_{r}\dot{\beta}^2
\end{equation}
where $(x_r, y_r)$ are the coordinates of the center of the rotor in the inertial frame of reference. 

The rear wheel or runner is not allowed slip in the transverse $(Y_b)$ direction, i.e.
  \begin{equation}\label{eq:Chaplygin_constraint}
 -\dot{x}\sin{\theta} + \dot{y}\cos{\theta} - b\dot{\theta} = 0
  \end{equation}
with Pfaffian one form being
 \begin{equation}\label{eq:pfaffian}
 -\sin{\theta}dx + \cos{\theta}dy - bd\theta = 0.
  \end{equation}

The nonholonomic constraint, \eqref{eq:Chaplygin_constraint} requires the use of the Lagrange multiplier method to derive the equations of motion. Such calculations for the Chaplygin sleigh can be found in, \cite{ft_jcnd_2017, tf_jcnd_2017} and these can be extended to the case of the Chaplygin sleigh with a passive internal rotor. As a first step the velocity of internal rotor is expressed in terms of its angular velocity and velocity of the sleigh to obtain a reduced Lagrangian

\begin{align}\label{eq:L}
\mathcal{L} = \frac{1}{2}m(\dot{x}^2+\dot{y}^2)+m_{r}a\dot{\beta}(\dot{y}\cos{\beta}-\dot{x}\sin{\beta}) \notag \\
+\frac{1}{2}I_c\dot{\theta}^2+\frac{1}{2}(I_{r}+m_{r}a^2)\dot{\beta}^2
\end{align}
where $m = m_r+m_c$.

The Euler-Lagrange equations are 

 \begin{equation} \label{eq:EL}
 \frac{d}{dt}\left(\frac{\partial \mathcal{L}}{\partial \dot{q_k}}\right)-\frac{\partial \mathcal{L}}{\partial q_k}= C_k\lambda
 \end{equation}
where $\lambda$ is the Lagrange multiplier and $C_{k}$ is the coefficient of the one form $dq_k$ in \eqref{eq:pfaffian}.

Straight forward calculations yield the Euler-Lagrange equations as
\begin{align} \label{eq:EL2}
m\ddot{x}-m_{r}\ddot{\beta}\sin{\beta}-m_{r}a\dot{\beta}^2\cos{\beta} &=& -\lambda\sin{\theta} \nonumber\\
m\ddot{y}+m_{r}\ddot{\beta}\cos{\beta}-m_{r}a\dot{\beta}^2\sin{\beta} &=& \lambda\cos{\theta} \nonumber\\
I_{c}\ddot{\theta}&=&-\lambda b  \nonumber\\
m_{r}a(\ddot{y}\cos{\beta}-\ddot{x}\sin{\beta}- \dot{y}\dot{\beta}\sin{\beta}-\dot{x}\dot{\beta}\cos{\beta})  \notag \\  +(I_r+m_{r}a^2)\ddot{\beta}&=&0.
\end{align}

In order to reduce the Euler-Lagrange equations and eliminate the constraint force, the velocities and accelerations of the tail may be expressed in terms of the longitudinal velocity of the wheel, $u$, and $\theta$,
 \begin{align}
 \dot{x}&= u \cos{\theta} - \dot{\theta} b\sin{\theta}  \label{eq:xdot}\\
 \dot{y}&= u \sin{\theta} + \dot{\theta} b\cos{\theta}  \label{eq:ydot}
 \end{align}
and
\begin{align}
\ddot{x} = \dot{u}\cos{\theta} - u\dot{\theta}\sin{\theta} -\Omega^2 b \cos{\theta} - \ddot{\theta}b \sin{\theta}\\
\ddot{y} = \dot{u}\sin{\theta} + u\dot{\theta}\cos{\theta}- \Omega^2 b \sin{\theta} + \ddot{\theta}b\cos{\theta}.
\end{align}

After substituting the above expressions into \eqref{eq:EL2} and eliminating $\lambda$ the following reduced equations are obtained.
\begin{equation}\label{eq:eom1}
\begin{bmatrix}
\textbf{I}& 0\\
\textbf{0} & 1
\end{bmatrix} 
\begin{bmatrix}
    \dot{u} \\
    \dot{\Omega} \\
    \dot{\omega}\\
\dot{\delta}
\end{bmatrix} 
=
\begin{bmatrix}
  m_cb\Omega^2+m_{r}a\omega^2\cos{\delta} \\
  -m_cbu\Omega-m_{r}ab\omega^2sin{\delta}  \\
    -m_{r}au\omega\cos{\delta}\\
\Omega- \omega
\end{bmatrix}
\end{equation}
where $\Omega = \dot{\theta}$, $\omega = \dot{\beta}$, $\delta = \theta - \beta$ is the angle made by the internal rotor with respect to the body $X_b$ axis and $\textbf{I}$ represents the locked inertia tensor,
\begin{equation}
\textbf{I}
=
\begin{bmatrix}
    m_c & 0 & m_{r}a\sin{\delta} \\
    0 & I_{c}+mb^2 & m_{r}ab\cos{\delta} \\
    m_{r}a\sin{\delta} & m_{r}ab\cos{\delta} & I_r+m_ra^2
\end{bmatrix}.
\end{equation}

We consider the special case where the mass, moment of inertia and length of the internal rotor are small compared to the corresponding parameters of the sleigh. We choose the small parameter

\begin{equation}
 \frac{m_r}{m} = \frac{a}{b} =\epsilon \ll 1.
\end{equation}
The reduced equations of motion are then defined by the dynamical system

\begin{align}
\dot{u}&= \frac{K^4b\Omega^2 + \epsilon (K^2-1)(b\Omega^2\cos^2{\delta}+u\Omega\sin{\delta}\cos{\delta})}{\epsilon(K^2-1)\cos^2{\delta}+K^4} \notag \\
&+\frac{b\omega^2 \epsilon(\epsilon(K^2-1)+K^2))\cos{\delta}}{\epsilon(K^2-1)\cos^2{\delta}+K^4} \label{eq:u}\\
\dot{\Omega}&= -\frac{K^2(\epsilon^2b\omega^2\sin{\delta}+u\Omega)}{b(\epsilon(K^2-1)\cos^2{\delta}+K^4)}  \label{eq:Omega}\\
\dot{\omega}&= -\frac{(K^2-1)\cos(\delta)(\epsilon^2b\omega^2\sin{\delta}+u\Omega)}{\epsilon b(\epsilon(K^2-1)\cos^2{\delta}+K^4)}  \label{eq:omega}\\
\dot{\delta} &= \Omega - \omega \label{eq:delta_dot}
\end{align}
where $K^2=1+\frac{I_r}{mb^2}$. This is a dynamical system that defines a flow on the manifold $\mathcal{M} = \mathbb{R}^3 \times S^1$. We will denote the flow map of this dynamical system by $\Phi_t^T :\mathcal{M} \mapsto \mathcal{M}$. Note that the only set of fixed points for eqs \eqref{eq:u}-\eqref{eq:delta_dot} is $\Omega=\omega=0$.

\section{Simulation Results}\label{sec:simulation}
\begin{figure}[!h]

 \begin{minipage}{0.48\hsize}
 	\begin{center}
 	 \includegraphics[width =\hsize]{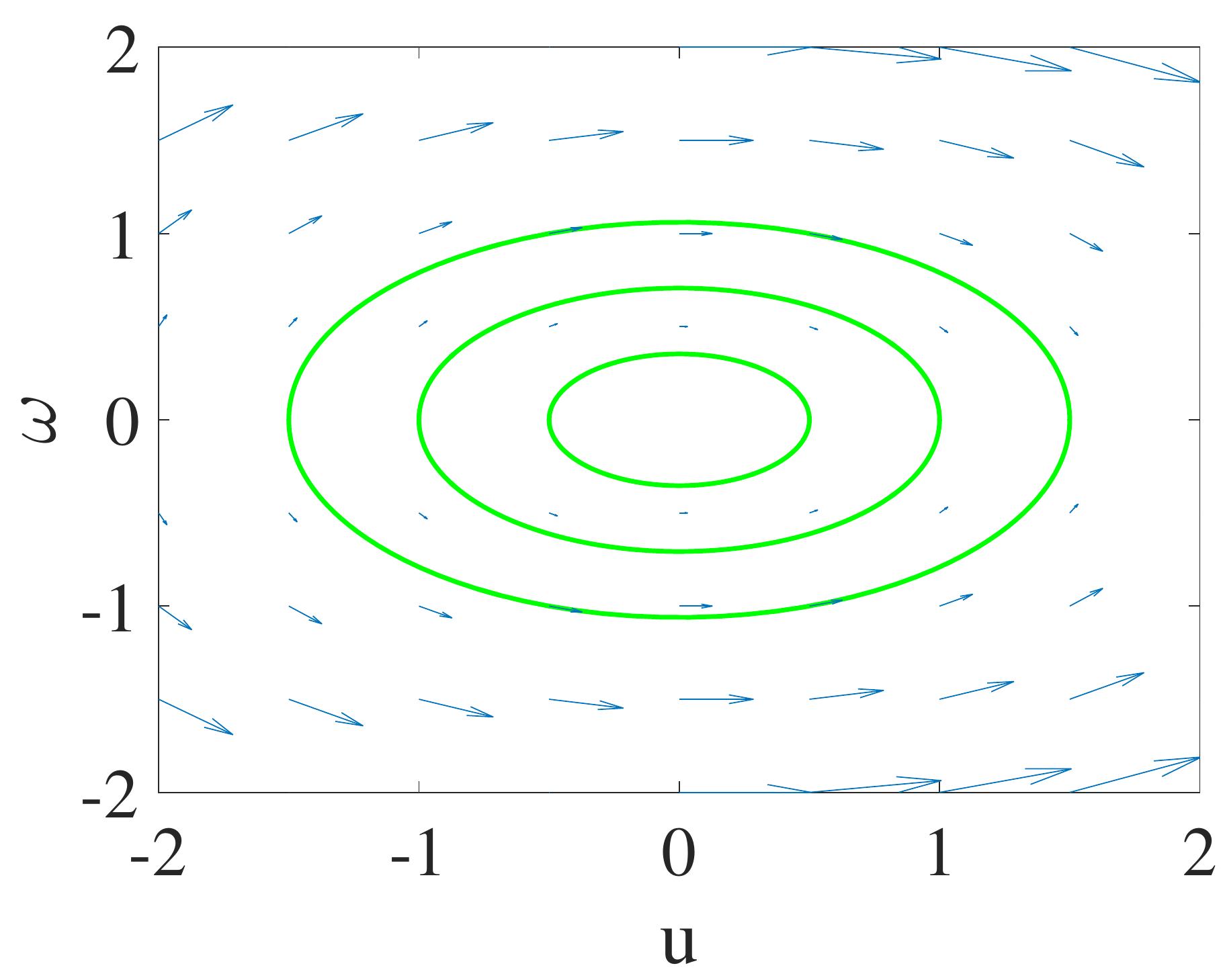}\\
 	   (a)
 	\end{center}
 \end{minipage}
 	 \begin{minipage}{0.48\hsize}
 	 	\begin{center}
 	 		\includegraphics[width =\hsize]{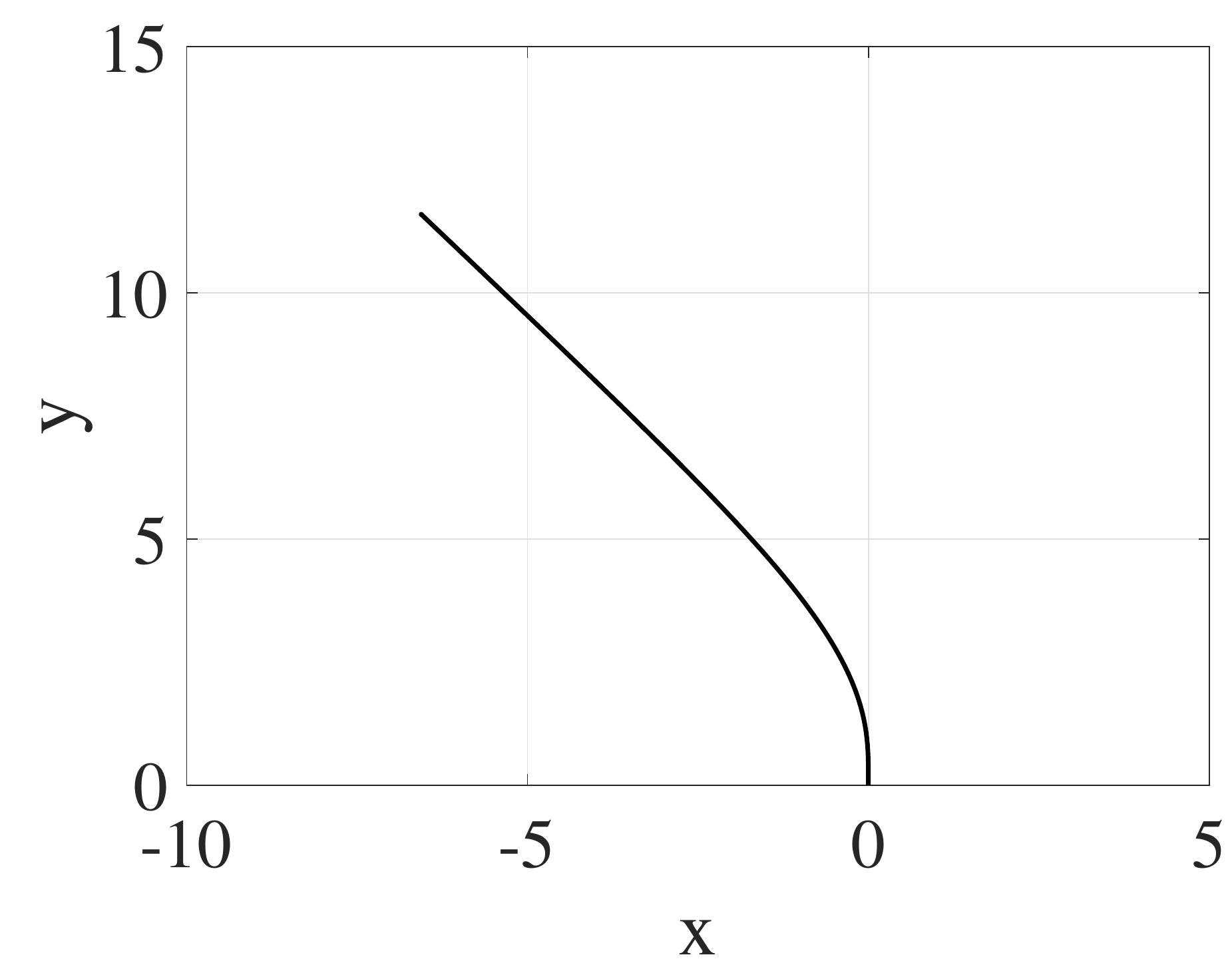}\\
 	 		(b)
 	 	\end{center}
 	 \end{minipage}
 	 
 \caption{A phase portrait of the Chaplygin sleigh is shown in a) with sample trajectories shown as solid curves. A generic trajectory of the sleigh is shown in b). Initial conditions are ($u(0)=0$,$\Omega(0)=1$).}\label{fig:phase_plot}
  \end{figure}

The equations of motion of the Chaplygin sleigh without the internal rotor, \cite{Neimark1972,ft_jcnd_2017, tf_jcnd_2017} can be obtained from \eqref{eq:u} and \eqref{eq:Omega} by setting $\epsilon = 0$,
\begin{align}
\dot{u}&= b\Omega^2 \label{eq:u1}\\
\dot{\Omega}&= -\frac{u\Omega}{bK^2}  \label{eq:Omega1}.
\end{align}
The phase portrait for this two dimensional dynamical system is shown in fig. \ref{fig:phase_plot}(a).  Since the right hand side of \eqref{eq:u1} is always non negative, $u$ is a non decreasing function.  In the reduced (velocity) state space defined by \eqref{eq:u1} and \eqref{eq:Omega1}, $\Omega= 0$ is a set of non isolated fixed points. The fixed points $(u<0, \Omega=0)$. The set $(u>0, \Omega=0$) is stable in the sense that any perturbation in the velocity from a fixed point in this set produces a trajectory that converges to the set $(u>0, \Omega =0)$, but not necessarily to the original fixed point. The kinetic energy of the sleigh is constant since the friction at point $P$ on the sleigh does not do any work, it can be shown that the kinetic energy is an integral of motion, \cite{Neimark1972, tf_jcnd_2017}. A decrease in the angular velocity of the sleigh leads to a corresponding increase in the longitudinal velocity of the sleigh.  Since the angular velocity, $\Omega$ converges to zero, the sleigh's trajectory in the physical plane converges to a straight line  that is unbounded, as shown in fig. \ref{fig:phase_plot}(b) for generic initial conditions.

\begin{figure}[!h]

 \begin{minipage}{0.48\hsize}
 	\begin{center}
 	 \includegraphics[width =\hsize]{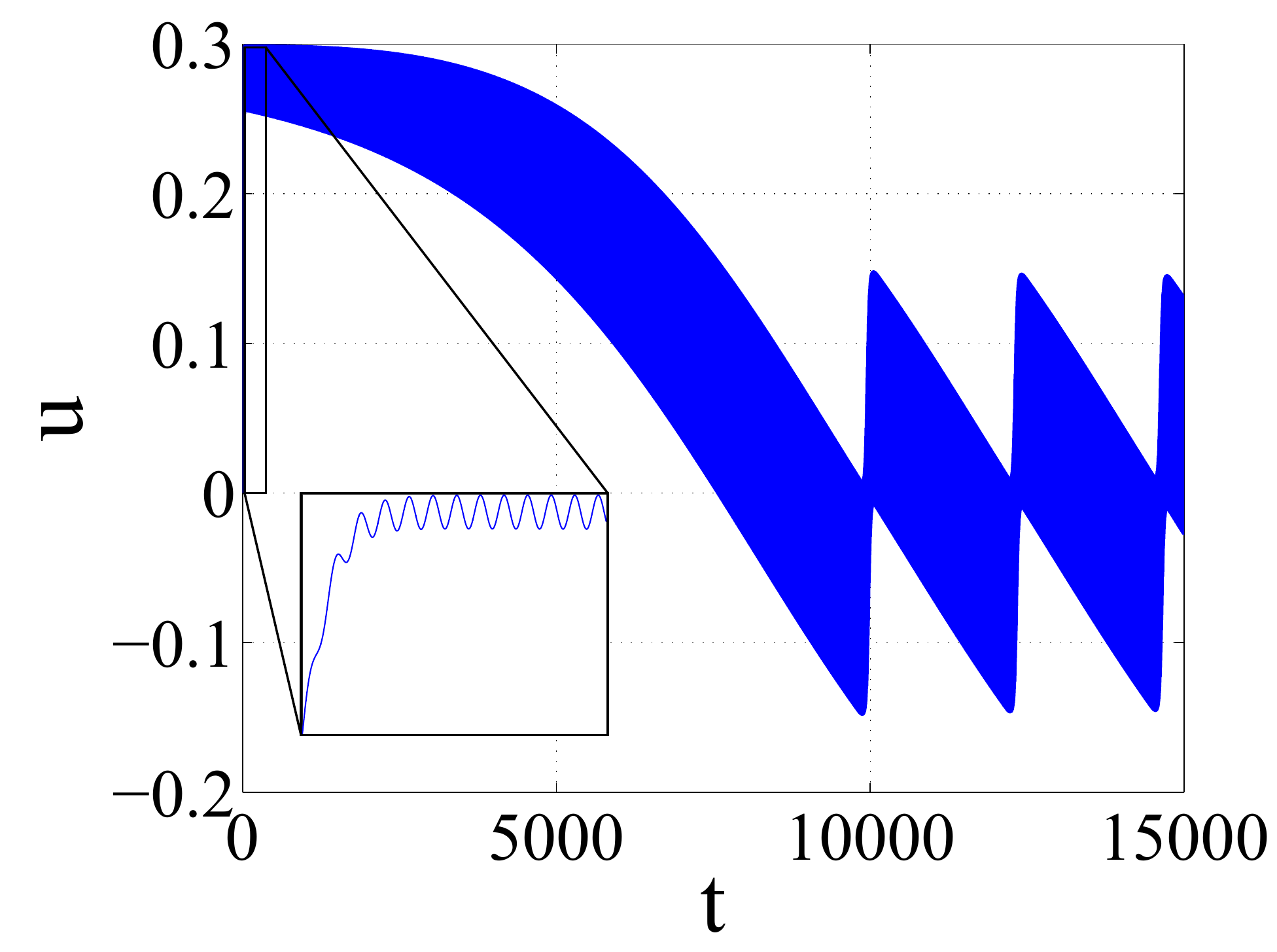}\\
 	   (a)
 	\end{center}
 \end{minipage}
 	 \begin{minipage}{0.48\hsize}
 	 	\begin{center}
 	 		\includegraphics[width =\hsize]{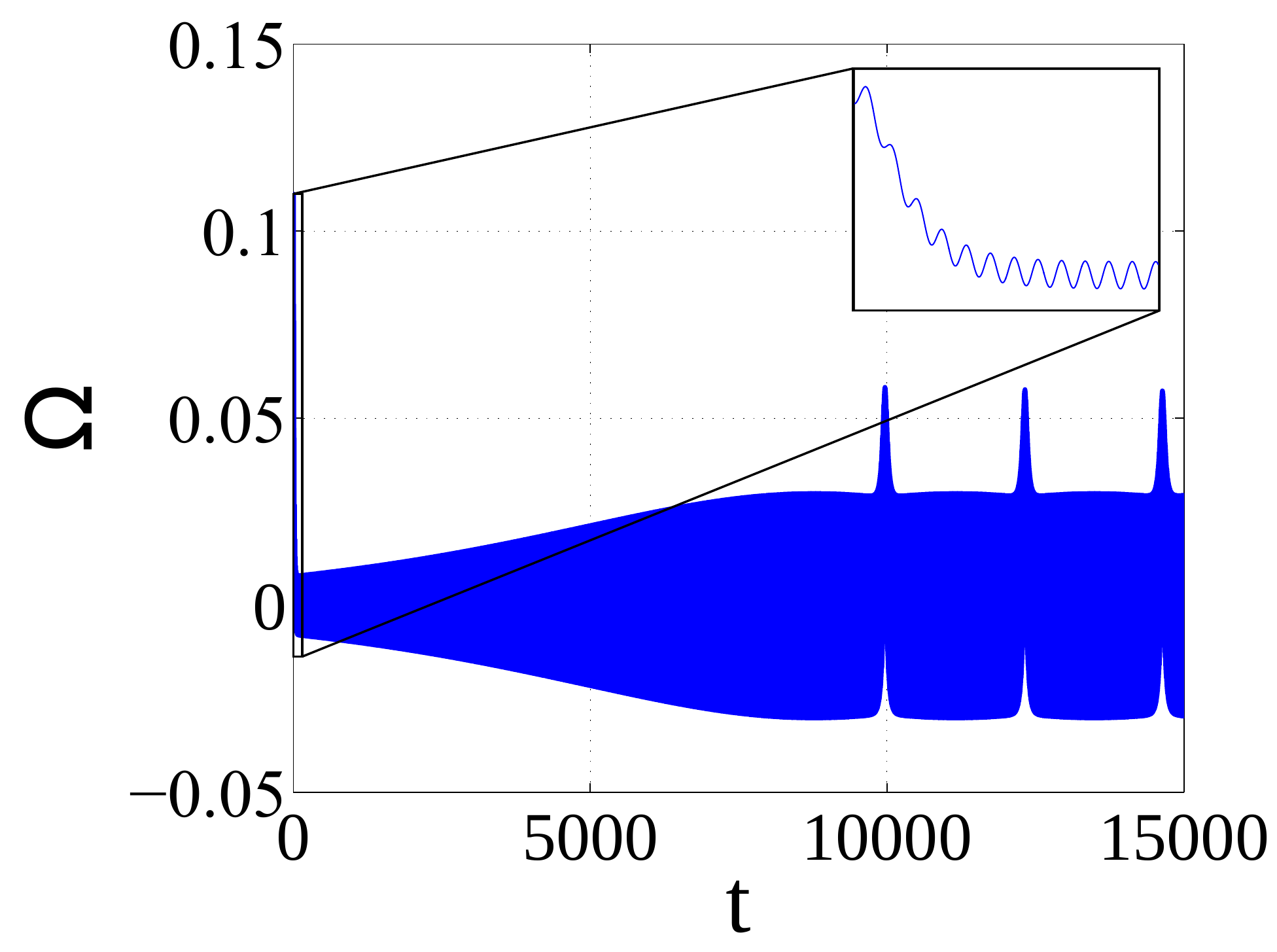}\\
 	 		(b)
 	 	\end{center}
 	 \end{minipage}
 	 
  \begin{minipage}{0.48\hsize}
  \begin{center}
  \includegraphics[width =\hsize]{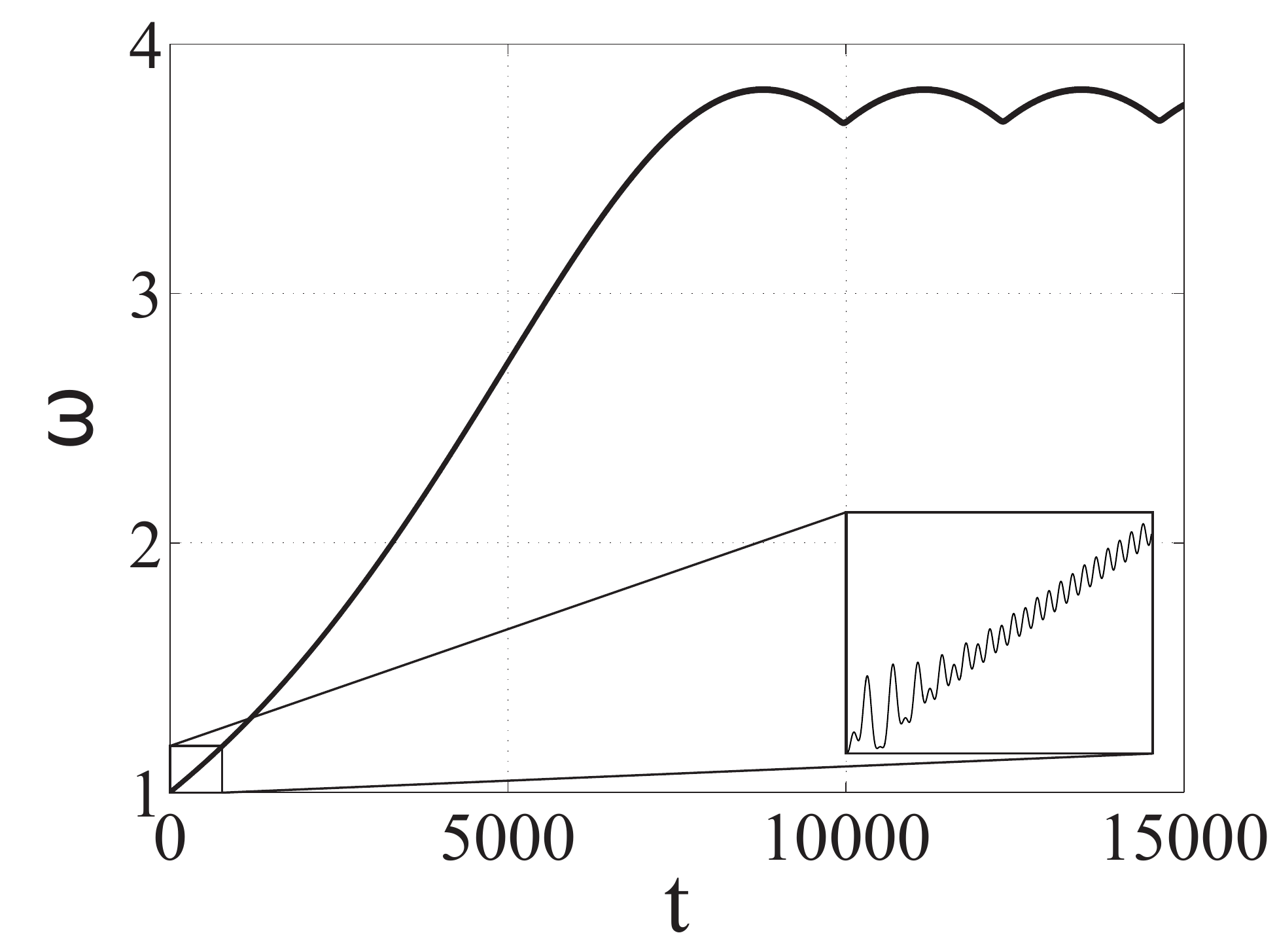}\\
  (c)
  \end{center}
  \end{minipage}
\begin{minipage}{0.48\hsize}
 	\begin{center}
 	 \includegraphics[width =\hsize]{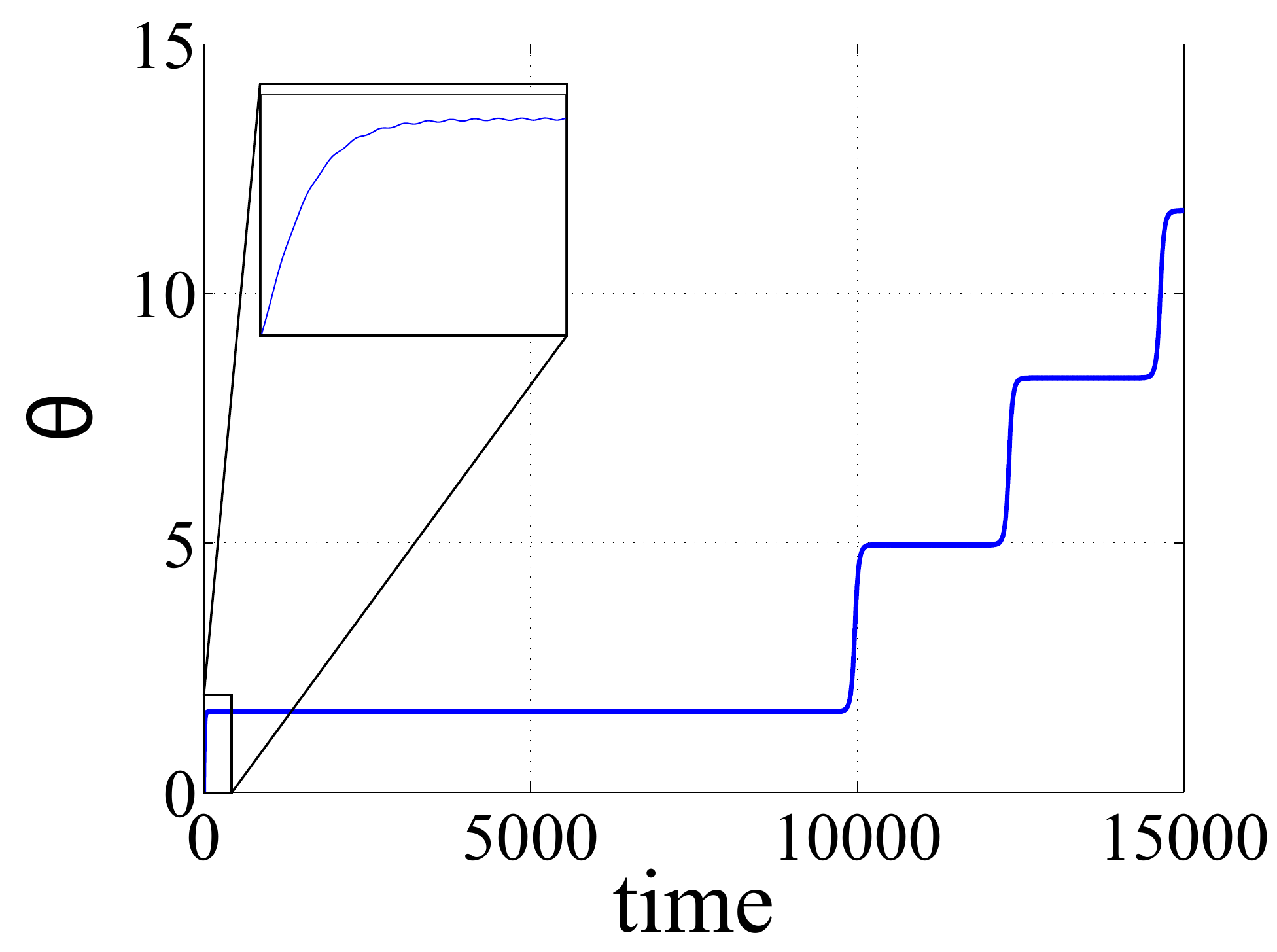}\\
 	   (d)
 	\end{center}
 \end{minipage}
 
  \caption{Simulation of the Chaplygin sleigh with a small internal rotor with $\epsilon = 0.1$ and initial conditions $u(0)=0$, $\Omega(0)=0.1$, $\omega(0)=1$. Energy is $E=0.1127$}\label{fig:two_link_long}
  \end{figure}

A direct simulation of the equations of motion of the sleigh with an internal rotor shows dramatically different dynamics. A sample simulation of \eqref{eq:u}, \eqref{eq:Omega} and \eqref{eq:omega} for generic initial conditions is shown in Fig. \ref{fig:two_link_long}.  The evolution of $u$ and $\Omega$ occurs with two distinct transient stages before convergence to a nearly periodic solution. These transient stages of dynamics are common to any initial condition, with variations in the time periods associated with these individual stages.

In the first transient phase highlighted in Fig.  \ref{fig:two_link_long}(a) and (b) it can be seen that for a short duration of time $(t < 100)$, the dynamics of the sleigh are nearly the same as that of a Chaplygin sleigh without the internal rotor. In this short interval the angular velocity of the sleigh becomes very small and the longitudinal velocity and $u$ reaches a nearly constant value. \\

In the second transient stage $(100<t<8,700)$,  $u(t)$ decays with oscillations and $\Omega(t)$ oscillates about zero but with an increasing amplitude. The angular velocity of the rotor increases, but with very small amplitude oscillations. At about $t=8700$ a steady state is reached and as will be discussed later, a trajectory converges to an attractor $\mathcal{A} \subset \mathcal{M}$.  From here on the longitudinal velocity of the sleigh has a nearly oscillatory behavior with two frequencies of oscillation. The longitudinal velocity is the sum of two periodic functions, one with a large time period of about $2200$ and zero mean and the other a periodic function with a very small time period. The angular velocity of the sleigh oscillates about zero, but with sudden spikes occuring at intervals of about $t=2200$. The angular velocity of the passive rotor show small oscillations around a non zero mean. The heading angle of the sleigh, $\theta$, is nearly piecewise constant with sudden jumps at time intervals of about $2200$.

\begin{figure}[!h]
\begin{minipage}{.48\hsize}
\begin{center}
\includegraphics[width =\hsize]{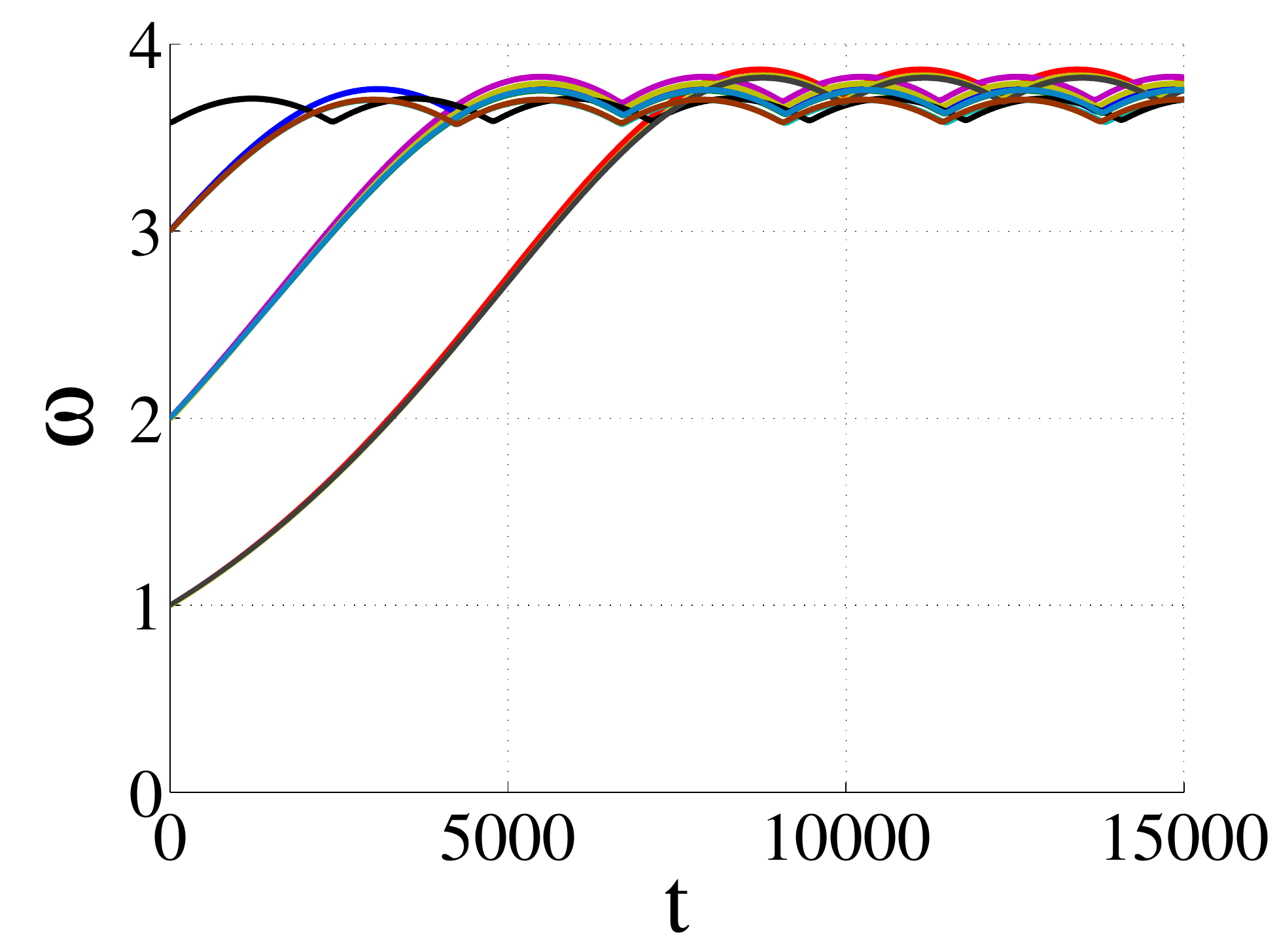}
\end{center}
\end{minipage}
\begin{minipage}{.48\hsize}
\begin{center}
\includegraphics[width =\hsize]{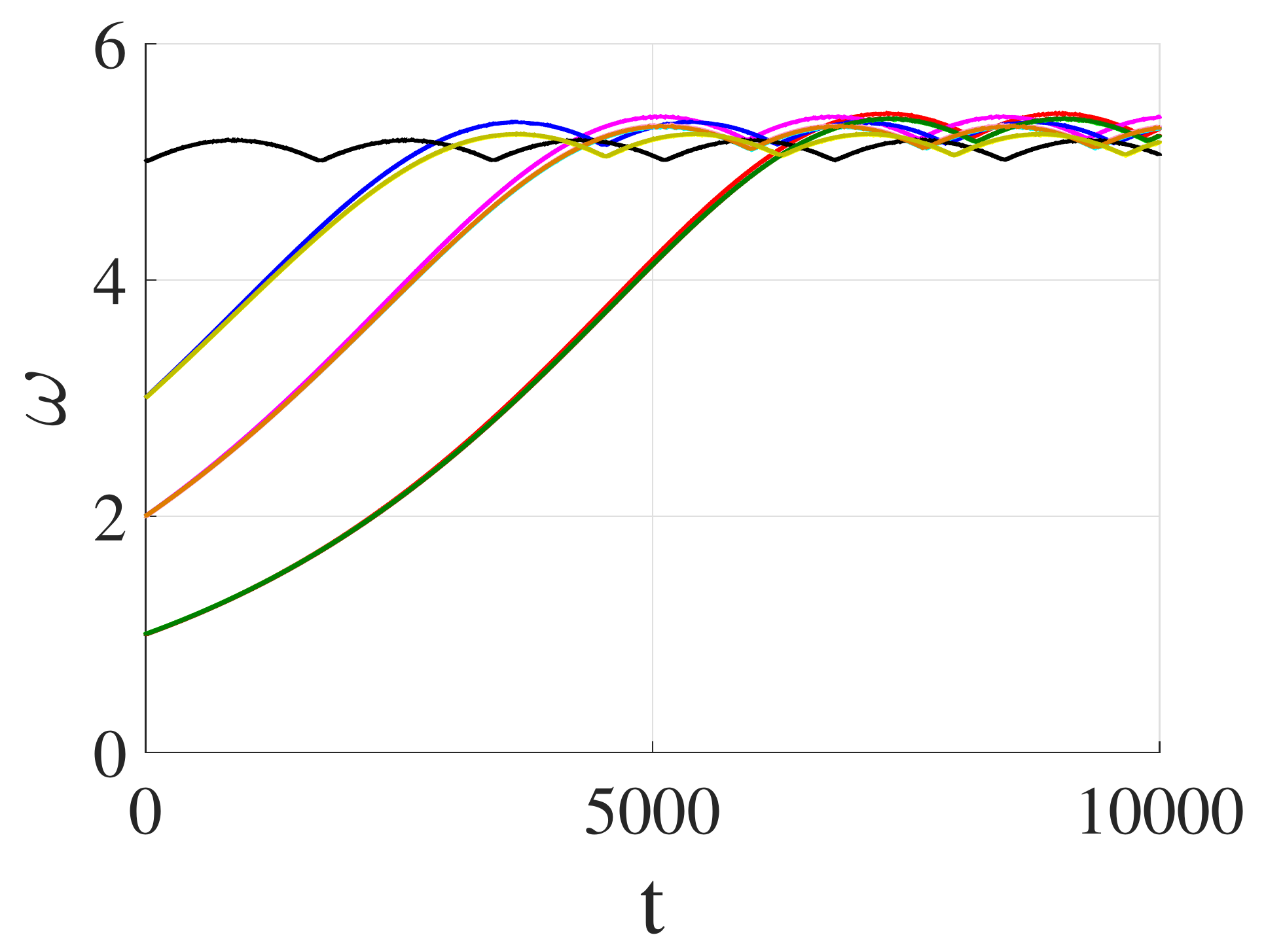}
\end{center}
\end{minipage}
\caption{Trajectory of the sleigh for different initial conditions on a level set of kinetic energy. (a) $E = 0.1184$ and (b) $E = 0.2313$.}\label{fig:multics}
\end{figure}

The transient dynamics of the sleigh and convergence to the attractor follow the same pattern  for any initial conditions. Initial conditions of the sleigh with distinct kinetic energies converge to distinct attractors. Conversely all initial conditions with the same kinetic energy converge to a unique attractor. This is shown most clearly in fig. \ref{fig:multics} where the angular velocity of the rotor for two sample sets of initial conditions on the same energy level is seen to converge to the same function for all initial conditions with the same energy. Furthermore, the initial transient dynamics are qualitatively the same for all the initial conditions shown in fig.\ref{fig:multics}.

\section{Transient dynamics of the sleigh and regular perturbation expansion}\label{sec:transient}

The first transient stage of the dynamics of the sleigh can be explained using a regular perturbation analysis of  \eqref{eq:u} -  \eqref{eq:delta_dot}.  Such regular perturbation analysis has been employed for the analysis of other nonholonomic systems such as the twist car, \cite{Or_PRE_2014, Or_PRS_2016, OR_JNLS_2017}.  We expand the states into a power series in $\epsilon$.
\begin{align}\label{eq:perturbation}
u(t) = & u_0(t) +\epsilon u_1(t) + \epsilon^2 u_2(t) + ...\\
\Omega(t) = & \Omega_0(t) +\epsilon \Omega_1(t) + \epsilon^2 \Omega_2(t) + ...\nonumber \\
\omega(t) = & \omega_0(t) +\epsilon \omega_1(t) + \epsilon^2 \omega_2(t) + ...\nonumber \\
\delta(t) = &\delta_0(t) + \epsilon\delta_1(t) + \epsilon^2\delta_2(t) + ...\nonumber
\end{align}
The right hand side of the preceding equations are expanded in a power series of $\epsilon$. We first note that $K^2-1=O(\epsilon)$. The denominators in each of the right hand side of the equations can then be expanded in a power series in $\epsilon$, 
\begin{align}
\frac{1}{\epsilon(K^2-1)\cos^2{\delta}+K^4} =& \frac{1}{K^4}(1-\epsilon\frac{K^2-1}{K^4}\cos^2{\delta}\nonumber\\
&+\epsilon^2\frac{(K^2-1)^2}{K^8}\cos^4{\delta}).\label{eq:den1}
\end{align}
We will also use the following,
\begin{align}\label{eq:trig_delta}
\sin{\delta} = & \sin{\delta_0} + (\epsilon\delta_1 + \epsilon^2\delta_2+...)\cos{\delta_0} + ...\\
\cos{\delta} = & \cos{\delta_0} - (\epsilon\delta_1 + \epsilon^2\delta_2+...)\sin{\delta_0} + ... \nonumber
\end{align}

Substituting  equations \eqref{eq:den1}, \eqref{eq:trig_delta} and the assumed power series expansion for $u$, $\Omega$ and $\omega$ into the \eqref{eq:u}, \eqref{eq:Omega} and \eqref{eq:omega} and equating the coefficients of the corresponding powers of $\epsilon$, one obtains the following equations for the three leading orders,

\begin{align}
\dot{u}_0 = & b\Omega_0^2 \label{eq:u0}\\
\dot{u}_1 = & 2b\Omega_0 \Omega_1  \label {eq:u_1}\\
\begin{split}
\dot{u}_2 = & \frac{1}{K^6}u_0\Omega_0 \sin{\delta_0} \cos{\delta_0}\\ 
 & +b\Omega_1^2+2b\Omega_0\Omega_2+bK^2\omega_0^2\cos{\delta_0}
\end{split}  \label {eq:u2}\\
\dot{\Omega}_0 = & -\frac{u_0\Omega_0}{bK^2} \label{eq:Omega0}\\
\dot{\Omega}_1 = & -\frac{u_1\Omega_0}{bK^2}-\frac{\Omega_1 u_0}{bK^2} \label {eq:Omega_1}\\
\begin{split}
\dot{\Omega}_2 = & \frac{1}{bK^6}u_0\Omega_0\cos^2{\delta_0} \\
 & -\frac{u_2\Omega_0}{bK^2}-\frac{u_1\Omega_1}{bK^2}-\frac{u_0\Omega_2}{bK^2} \\
 & -\frac{1}{K^2}\omega_0^2\sin{\delta_0} \label{eq:Omega2}
\end{split}\\
\dot{\omega}_0 = & -\frac{u_0\Omega_0}{bK^4}\cos{\delta_0} \label{eq:omega0}\\
\dot{\omega}_1 = & -\frac{1}{bK^4}(u_1\Omega_0+u_0\Omega_1)\cos{\delta_0} \label {eq:omega_1}\\
\begin{split}
\dot{\omega}_2 = & \Big(\frac{1}{bK^8}u_0\Omega_0\cos^2{\delta_0} -\frac{u_2\Omega_0}{bK^4}-\frac{u_1\Omega_1}{bK^4}-\frac{u_0\Omega_2}{bK^4} \\
 & -\frac{1}{K^4}\omega_0^2\sin{\delta_0}\Big) \cos{\delta_0}+\frac{1}{K^4b}u_0 \Omega_0\delta_2\sin(\delta_0) \label{eq:omega2}.
\end{split}
\end{align}

\begin{figure}[!h]
 \begin{minipage}{0.31\hsize}
 	\begin{center}
 	 \includegraphics[width =\hsize]{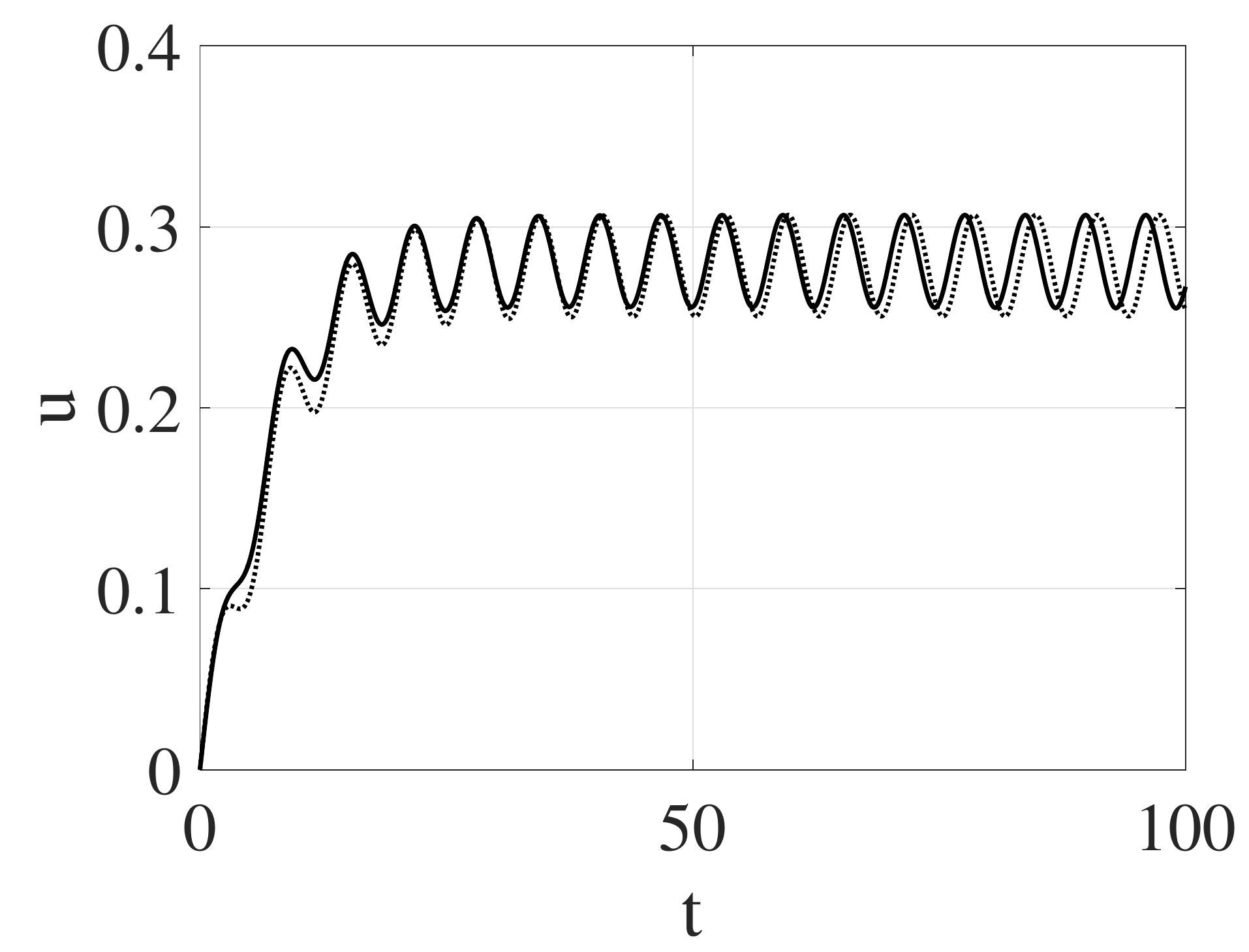}\\
 	   (a)
 	\end{center}
 \end{minipage}
 	 \begin{minipage}{0.31\hsize}
 	 	\begin{center}
 	 		\includegraphics[width =\hsize]{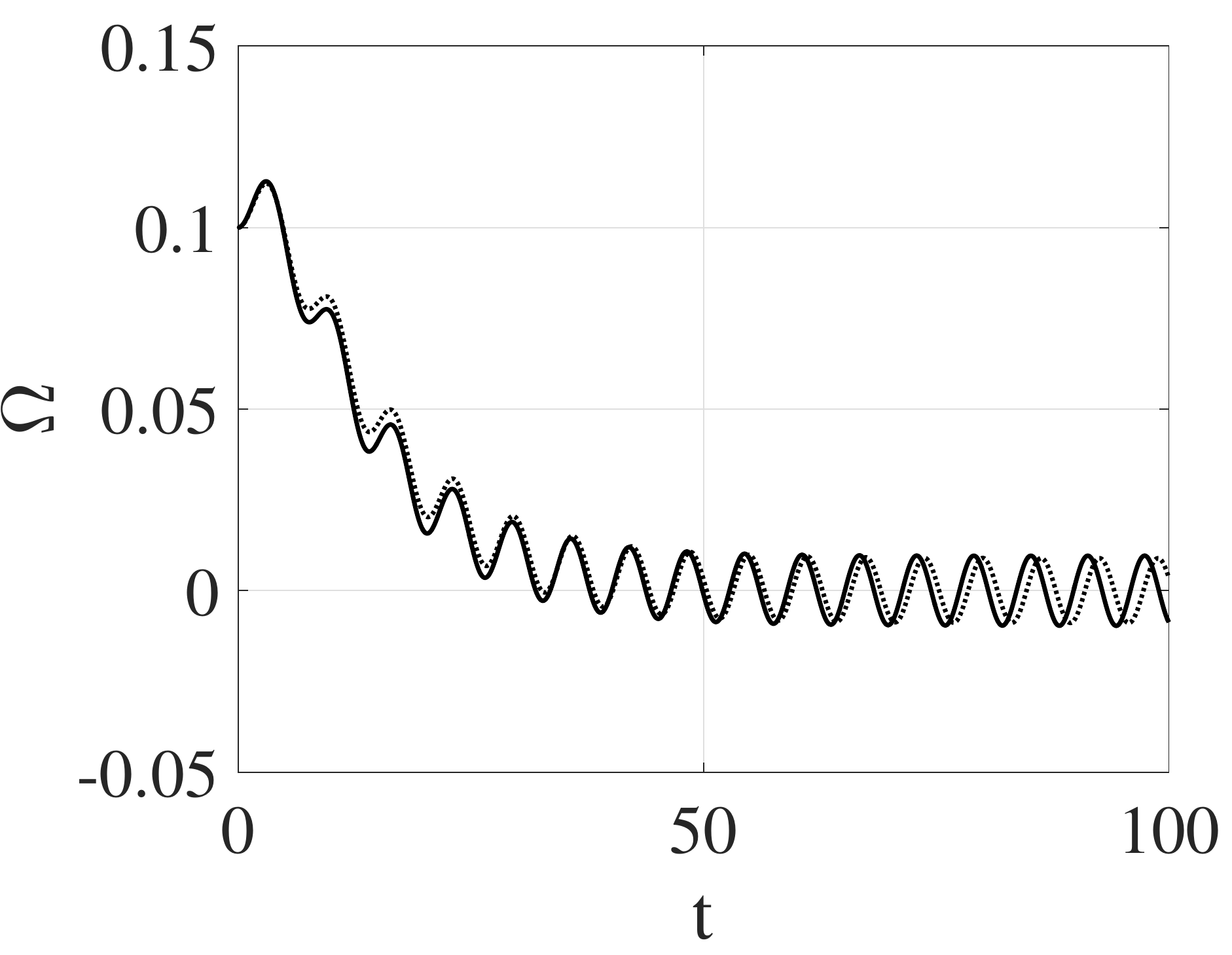}\\
 	 		(b)
 	 	\end{center}
 	 \end{minipage}
  \begin{minipage}{0.31\hsize}
  \begin{center}
  \includegraphics[width =\hsize]{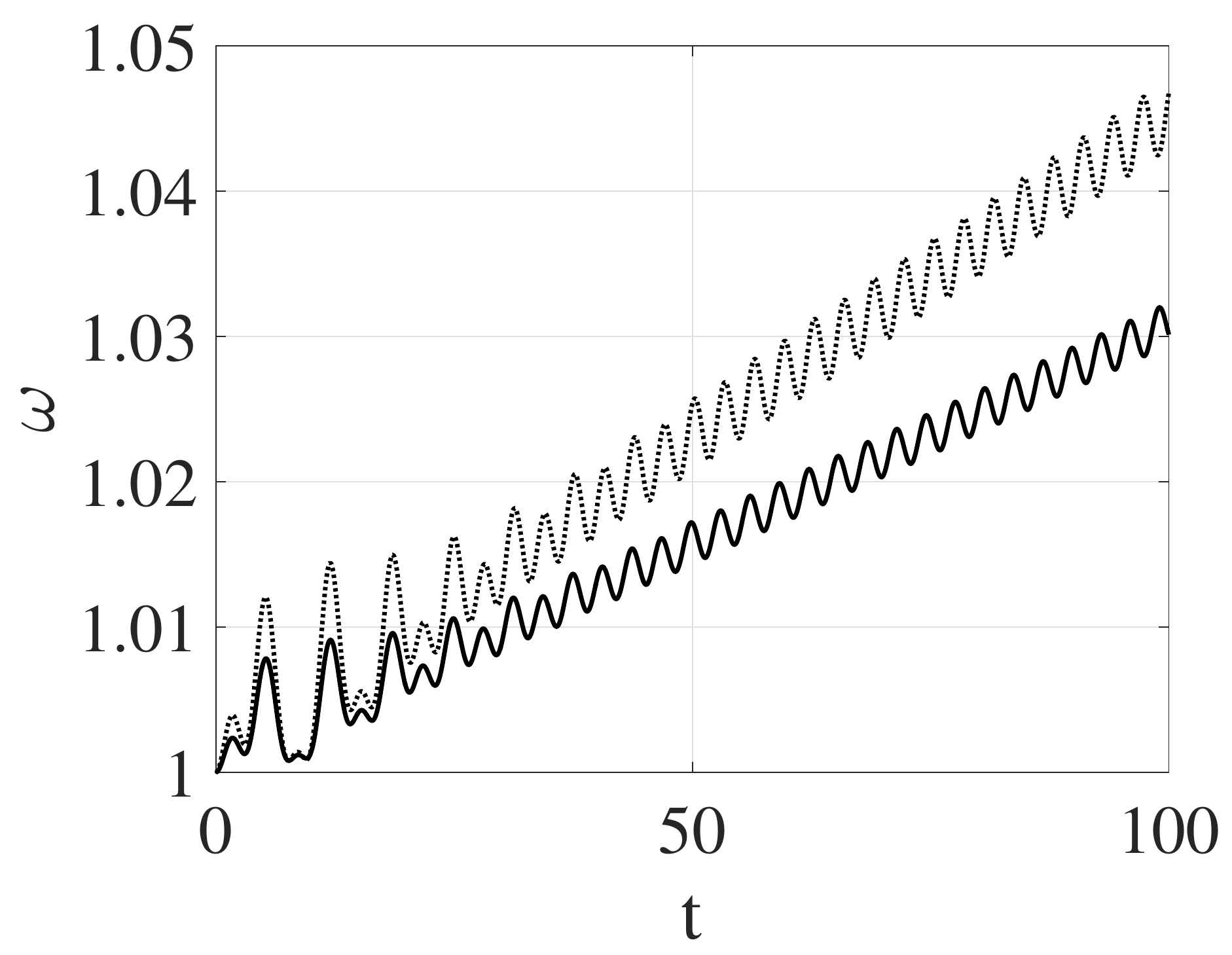}\\
  (c)
  \end{center}
  \end{minipage}
    \caption{Simulation of the Chaplygin sleigh with a small internal rotor for initial conditions $u(0)=0$, $\Omega(0)=.1$, $\omega(0)=1$ ($u_0(0)=0$, $\Omega_0(0)=.1$, $\omega_0(0)=1$ and other initial conditions are zero for the regular expansion). The solution due to the perturbation expansion is shown as a dashed line and the exact solution as a solid line.}\label{fig:two_link_short}
 \end{figure}

The validity of the regular perturbation expansion for short time periods  is borne from the close match between the solution $u_0(t) + \epsilon u_1(t) + \epsilon^2 u_2(t)$ with the solution of \eqref{eq:u} as shown in Fig. \ref{fig:two_link_short}(a). A similar comparison for the angular velocities is shown in   Fig. \ref{fig:two_link_short}(b) and (c). The initial conditions for a direct numerical simulation of \eqref{eq:u}, \eqref{eq:Omega} and \eqref{eq:omega} are $(u(0) = 0, \Omega(0) = 0.1, \omega(0) = 1)$. Since the perturbation expansion, \eqref{eq:perturbation} is valid for any arbitrarily small value of $\epsilon$,  the initial conditions for simulation of the perturbation expansion equations are $(u_0(0) =0, u_1(0) = 0, u_2(0)=0)$, $(\Omega_0(0) =0.1, \Omega_1(0) = 0, \Omega_2(0)=0)$ and  $(\omega_0(0) =1, \omega_1(0) = 0, \omega_2(0)=0)$. 

 In Fig. \ref{fig:two_link_short} (a) and (b) we see that the longitudinal velocity of the sleigh increases from zero and begins to show small oscillations with a non zero mean. The angular velocity of the sleigh, $\Omega$, decays to zero with small oscillations. Except for the oscillations in the growth of $u$ and the decay of $\Omega$, the evolution of these velocities is similar to those of the regular Chaplygin sleigh. But more significantly there is a transfer of kinetic energy to the motion of the internal rotor, with $\omega(t)$ experiencing slow growth with oscillations. The oscillations in $u(t)$, $\Omega(t)$ and $\omega(t)$ can be explained by the  regular perturbation expansion equations. The error between the perturbation solution for $\omega$ and a direct numerical simulation grows faster and even on a time scale of $100$s the error is about $2\%$. This can be expected since the angular velocity $\omega(t)$ can have secular growth as seen in fig. \ref{fig:two_link_long}

\begin{figure}[!h]
\begin{minipage}{0.48\hsize}
 	\begin{center}
 	 \includegraphics[width =\hsize]{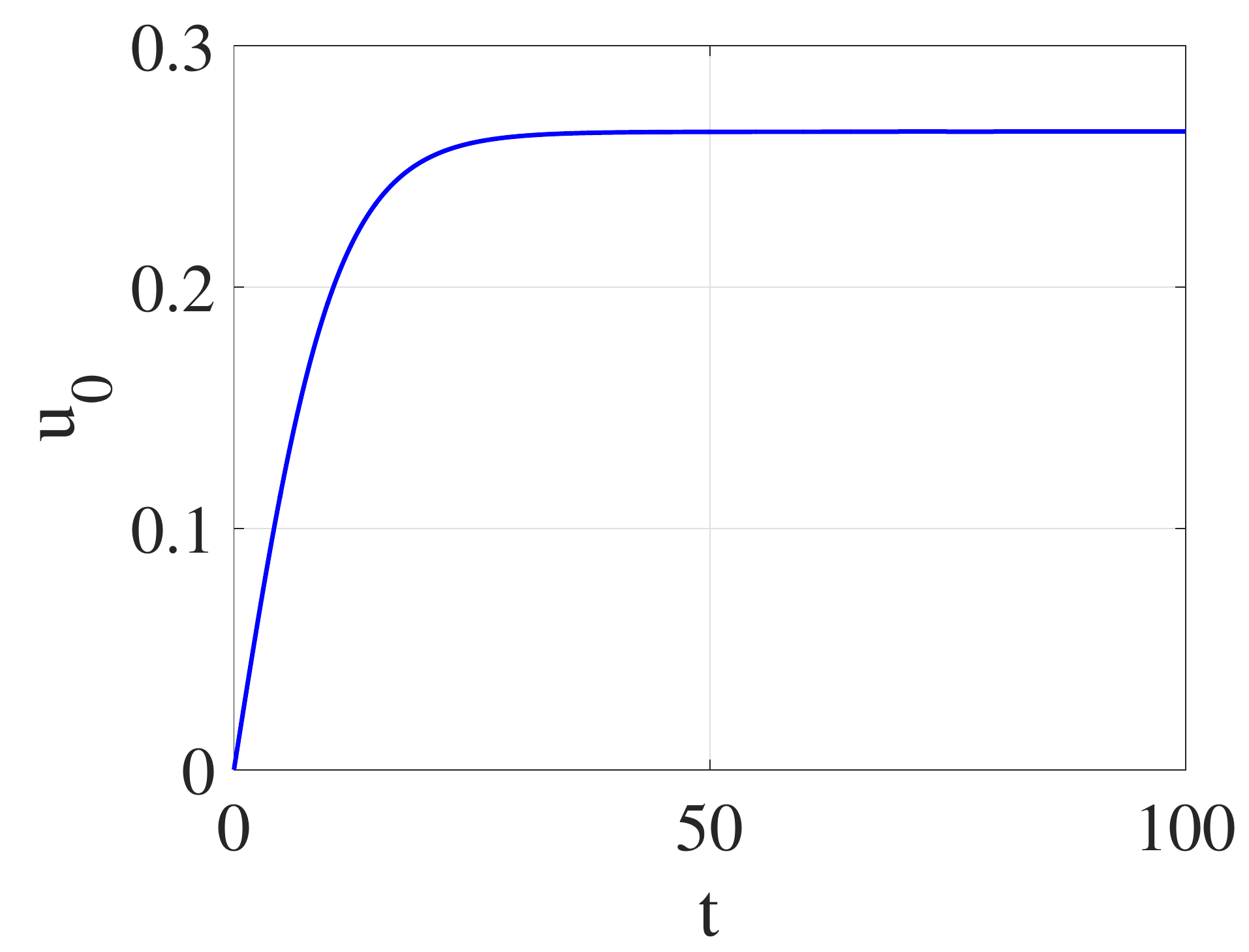}\\
 	   (a)
 	\end{center}
 \end{minipage}
 \begin{minipage}{0.48\hsize}
 	\begin{center}
 	 \includegraphics[width =\hsize]{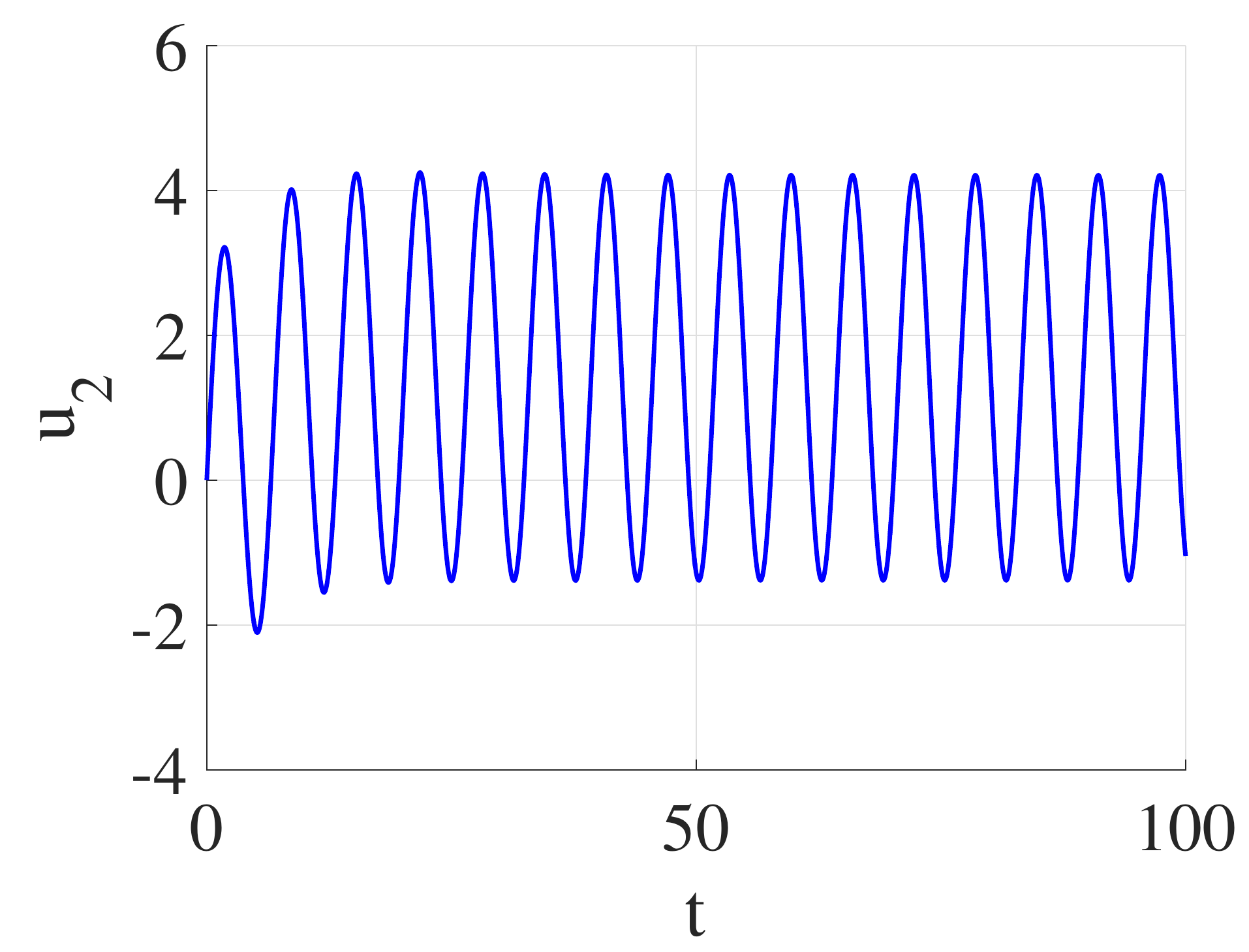}\\
 	   (b)
 	\end{center}
 \end{minipage}
 
 	 \begin{minipage}{0.48\hsize}
 	 	\begin{center}
 	 		\includegraphics[width =\hsize]{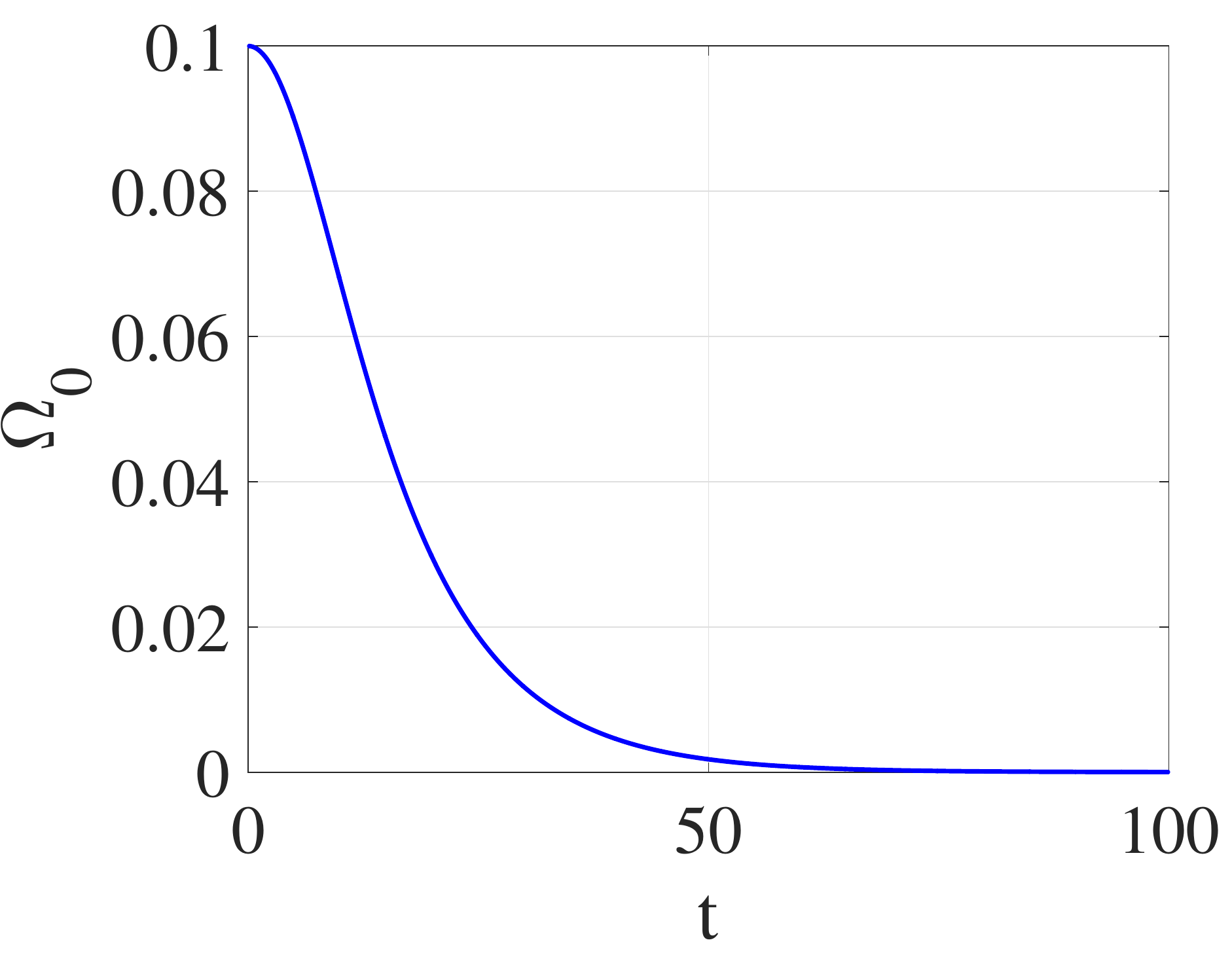}\\
 	 		(c)
 	 	\end{center}
 	 \end{minipage}
  \begin{minipage}{0.48\hsize}
  \begin{center}
  \includegraphics[width =\hsize]{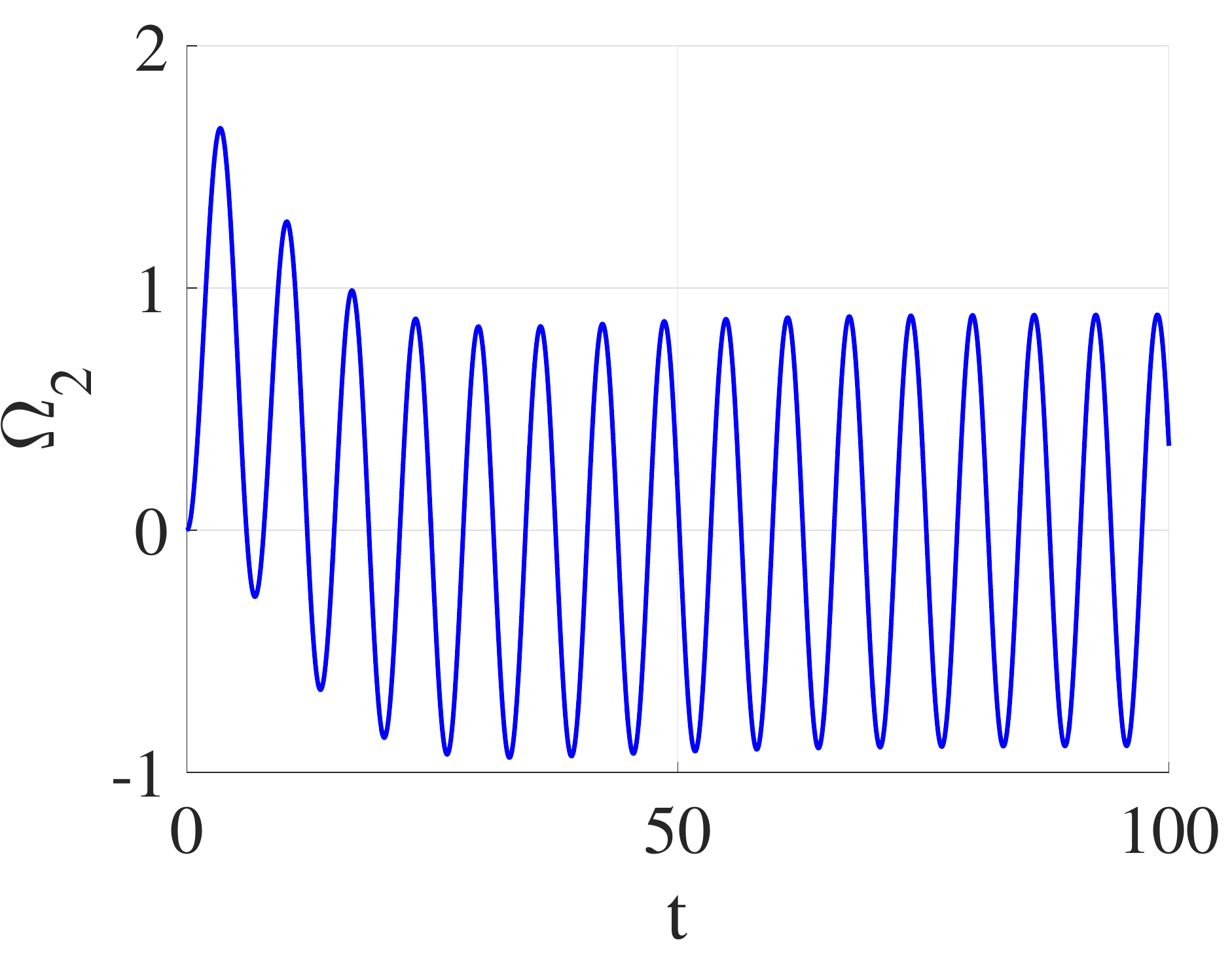}\\
  (d)
  \end{center}
  \end{minipage}
  
   	 \begin{minipage}{0.48\hsize}
 	 	\begin{center}
 	 		\includegraphics[width =\hsize]{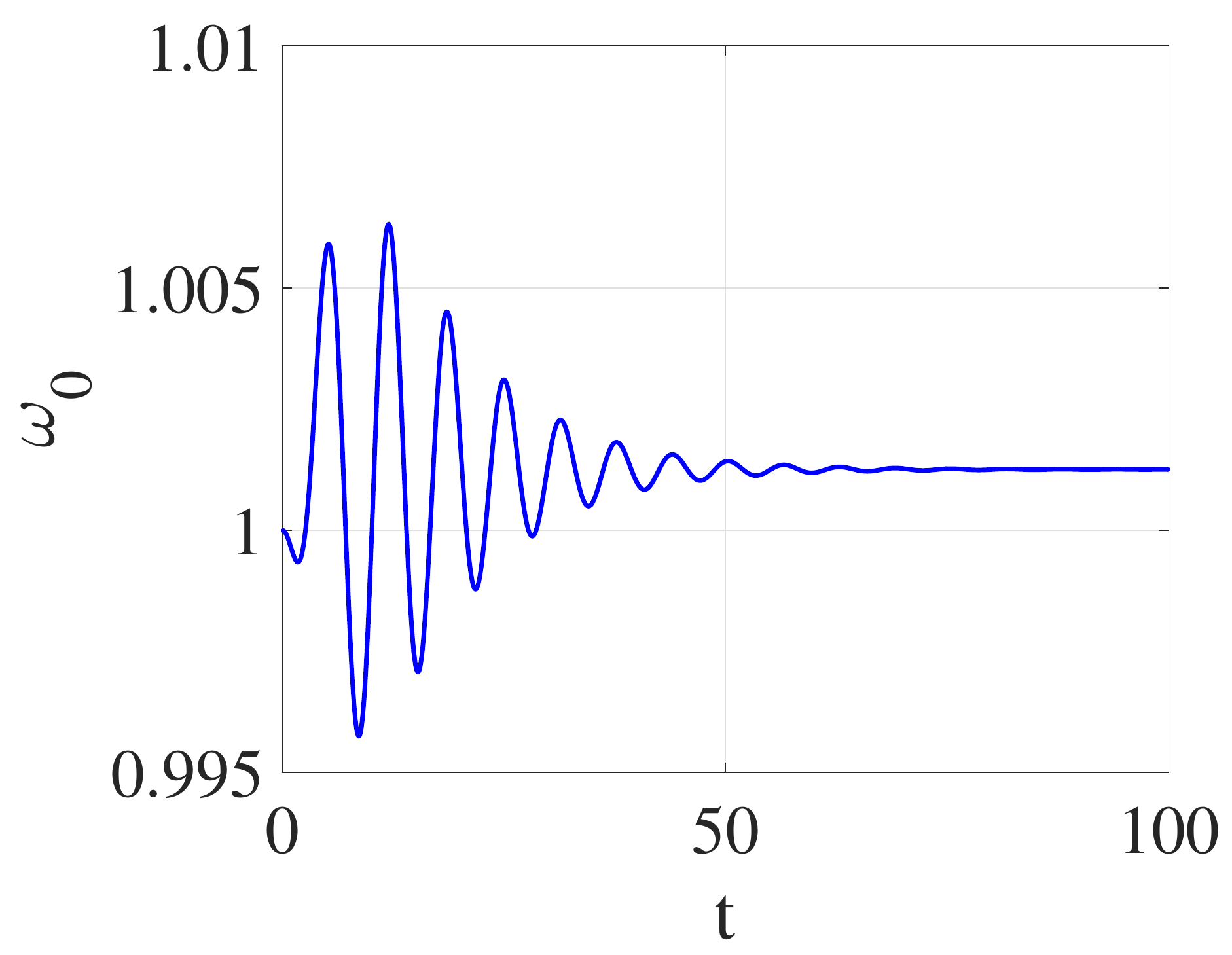}\\
 	 		(e)
 	 	\end{center}
 	 \end{minipage}
  \begin{minipage}{0.48\hsize}
  \begin{center}
  \includegraphics[width =\hsize]{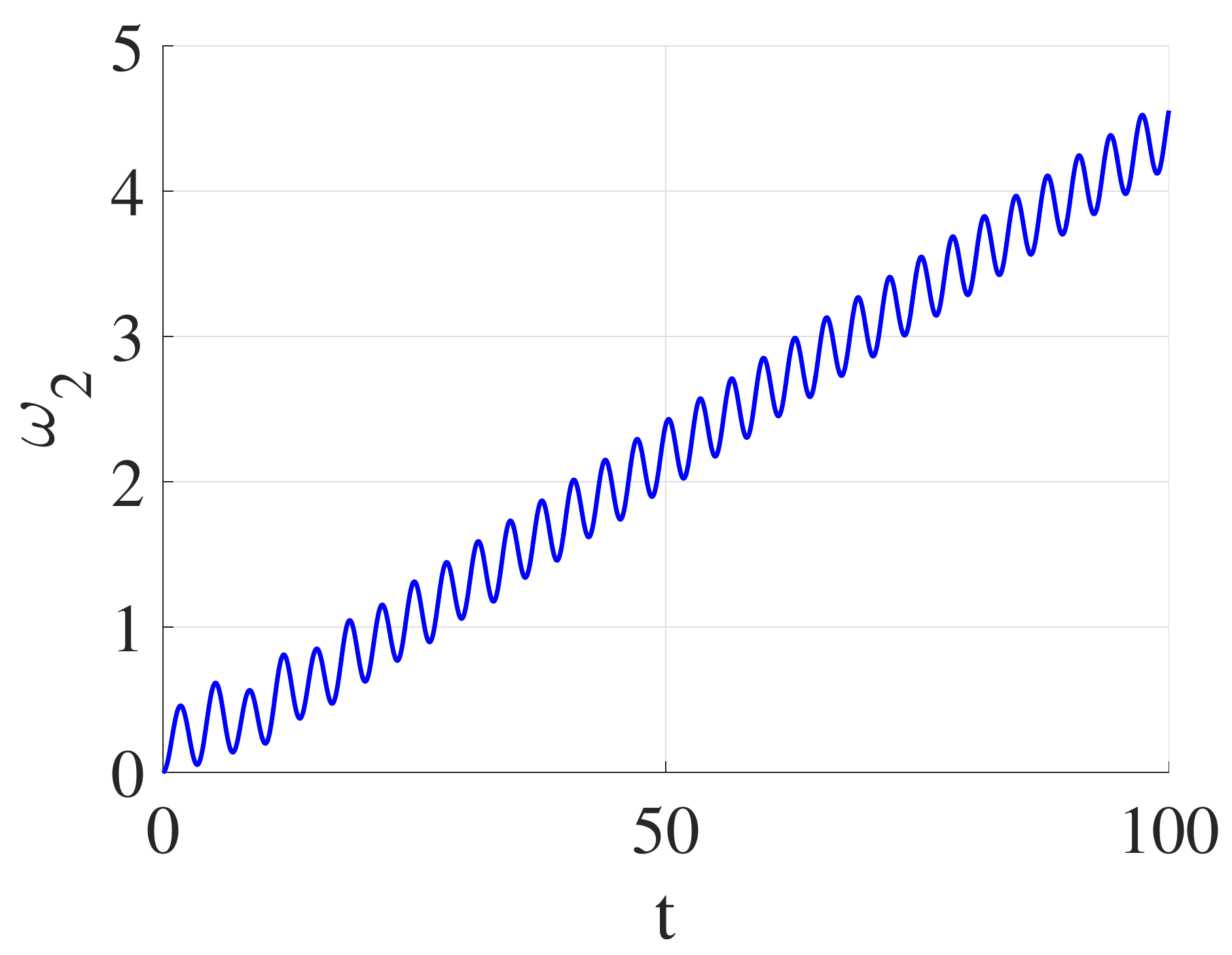}\\
  (f)
  \end{center}
  \end{minipage}
    \caption{Order $\epsilon^0$ and order $\epsilon^2$ solutions for the velocities of the sleigh with initial conditions $u_0=0$, $\Omega_0=.1$, $\omega_0=1$, $\delta_0=0$ and all other initial conditions zero. Order one $\epsilon^1$ solutions remain zero for all time.}\label{fig:pert_terms}
 \end{figure}

Equations \eqref{eq:u0} and \eqref{eq:Omega0} represent the leading order equations for the evolution of the velocity of the sleigh. These equations are the same as those that describe the motion of the Chaplygin sleigh without an internal rotor, \eqref{eq:u1}, \eqref{eq:Omega1}.  The leading order solutions $u_0$ and $\Omega_0$ are shown in Fig. \ref{fig:pert_terms}(a) and (c). The $O(\epsilon^0)$ solutions exhibit the behavior of a regular Chaplygin sleigh without an internal rotor.

The right hand sides of equations \eqref{eq:u_1} and \eqref{eq:Omega_1} are zero since the initial conditions $(u_1(0) =0, \Omega_1(0) = 0, \omega_1(0) =0)$ are the fixed points of the $O(\epsilon)$ equations. Therefore the $O(\epsilon)$ solution is always zero.

The first term on the right hand side of \eqref{eq:omega0} decays to zero since $\Omega_0$ decays to zero. Therefore $\omega_0$ converges to a constant value as shown in Fig. \ref{fig:pert_terms}(e). The first term on the right hand side of \eqref{eq:u2} decays to zero since $\Omega_0$ decays to zero. The second term $b\Omega_1^2$ is zero and the third term $2b\Omega_0 \Omega_2$ decays to zero. Therefore the equation \eqref{eq:u2} can be approximated as
\begin{equation}\label{eq:u2_approx}
\dot{u}_2 \approx bK^2\omega_0^2\cos{\delta_0}.
\end{equation}
 We next make a series of approximations  for the relative angle $\delta$ by first noting that
\begin{align}\label{eq:delta}
\delta(t) =& \delta(0) + \int_{0}^{t}(\Omega - \omega)dt  \notag\\
=&  \delta(0) +\int_{0}^{t}(\Omega_0 - \omega_0)dt  + \epsilon^2\int_{0}^{t}(\Omega_2 - \omega_2)dt + ... .
\end{align}
Since $\Omega_0$ decays to zero rapidly and $\omega_0$ reaches a constant value rapidly, we will make the approximation
\begin{equation}
\delta_0 \approx \delta(0) -\omega_0 t
\end{equation}
Assuming $\delta(0)=0$, it is clear from \eqref{eq:u2_approx}, $u_2$ shows oscillatory behavior, with the oscillations becoming nearly periodic after a short transient, as shown in Fig. \ref{fig:pert_terms}(b).

A similar argument can be made for the right hand side of \eqref{eq:Omega2}. First we set $u_1(t)=0$, $\Omega_1(t) = 0$ and let $\Omega_0\to 0$ to obtain

\begin{equation}\label{eq:Omega2_approx}
\dot{\Omega}_2 \approx -\frac{u_0\Omega_2}{bK^2}-\frac{1}{K^2}\omega_0^2\sin{\delta_0}.
\end{equation}
The stable solutions to the leading order equations are such that $u_0>0$. Therefore first term on the right hand side of \eqref{eq:Omega2_approx} causes a decay of $\Omega_2 \rightarrow 0$. The second term causes periodic oscillations around zero, as shown in Fig. \ref{fig:pert_terms}(f). Equation \eqref{eq:Omega2_approx} is a linear differential equation with a periodic forcing, the steady state solution for which is also periodic with the same frequency as the forcing frequency,
\begin{equation}\label{eq:Omega2_sol}
\Omega_2 = -\left(\frac{1}{\sqrt{\omega_0^2 + \frac{u_0^2}{b^2K^4}}}\right) \frac{\omega_0^2}{K^2}\sin{(\delta + \phi_1)}
\end{equation}
where $\phi_1 = tan^{-1}(\frac{-\omega_0bK^2}{u_0}) = 0.4705 \pi \approx \pi/2$. The amplitude of the steady solution of $\Omega_2$ is $0.892$.

 The evolution of $u_2$ and $\Omega_2$ by the simulation of \eqref{eq:u0}-\eqref{eq:omega2} is shown in Fig. \ref{fig:pert_terms}(b) and (d). The two velocities converge to oscillatory solutions with a time period of $T_1 = 6.276$ that is nearly equal to $\frac{2\pi}{\omega_0} = 6.277$, which bears out the validity of the series of approximations we made leading to \eqref{eq:Omega2_sol}.

The steady state behavior of $\omega_2$ , seen in fig. \ref{fig:pert_terms}(f) is a periodic function with time period $T_2 = 3.138$ which is half that of $T_1$. This can be understood by examining the several terms on the right hand side of \eqref{eq:omega2}. Since $\Omega_0 \rightarrow 0$ and $u_1 =0$ and $\Omega_1=0$, the steady state evolution of $\omega_2$ is approximately governed by the equation
\begin{align*}
\dot{\omega}_2 =& -\frac{u_0\Omega_2}{bK^2}\cos{\delta_0}-\frac{1}{K^2}\omega_0^2\sin{\delta_0}\cos{\delta_0} \notag \\
\approx & \frac{u_0A}{bK^2}\cos^2{\delta_0}-\frac{1}{2K^2}\omega_0^2\sin{2\delta_0}.
\end{align*}
where we substituted for the steady state solution of $\Omega_2$ from \eqref{eq:Omega2_sol} with the further approximation that the phase angle $\phi_1$, in \eqref{eq:Omega2_sol} is approximately $\pi/2$. It is straightforward to show that 

\begin{align}\label{eq:w2_approx}
\dot{\omega}_2 =&\frac{u_0A}{2bK^2} + \frac{u_0A}{2bK^2} \cos{2\delta_0} -\frac{\omega_0^2}{2K^2}\sin{2\delta_0} \notag \\ 
= & \frac{u_0A}{2bK^2}  - B \sin{(2\delta_0 - \phi_2)} \notag \\ 
\approx & \frac{u_0A}{2bK^2}  + B \sin{(2\omega_0t + \phi_2)} 
\end{align}
where 
\[
B = \sqrt{\left(\frac{u_0A}{2bK^2}\right)^2 + \left(\frac{\omega_0^2}{2K^2}\right)^2}
\]
and $\phi_2 = tan^{-1}\left(\frac{u_0A}{b\omega_0^2}\right)$. The constant term and the periodic term on the right hand side of \eqref{eq:w2_approx} produce respectively a linear growth in $\omega_2$ and an oscillatory response with time period $T_2 = \frac{\pi}{\omega_0}$. Using the previously obtained values of $\omega_0$ and $u_0$ we find that $T_2 = 3.135$, $B = 0.238$  and the average value of $\Omega_2$ grows linearly at a rate of $0.0415$. A direct simulation of equation \eqref{eq:omega2} shown in fig. \ref{fig:pert_terms}(f) show that time period $T_2 = 3.138$, $B = 0.241$ and the linear growth rate is $0.0445$ which are in very good agreement with the values obtained through the analytical approximations. Note that the growth rate is approximate and any discrepancy from the true growth rate causes the error in $\omega$ to accumulate. This is why in Fig. \ref{fig:two_link_short} (c) we see the approximate and actual solutions slowly growing apart.

As  fig. \ref{fig:two_link_long} (a) shows the longitudinal velocity decays with oscillations after about $t>100$ until about $t= 8,700$. In the same time period the angular velocity of the sleigh oscillates about zero with the amplitude steadily increasing until it reaches a nearly steady value as seen in fig. \ref{fig:two_link_long} (b). This second transient phase is generic to any initial conditions with positive longitudinal velocity, although the time period associated with this transient behavior changes with the initial conditions. 

This second transient phase cannot be explained through a perturbation analysis. The third order perturbation solutions, $u_3$, $\Omega_3$ and $\omega_3$ turn out to be zero. The equations for the fourth order variables contain many secular terms, leading to unbounded solutions.  It is however easy to see that the velocities, $u$, $\Omega$ and $\omega$ should remain bounded since the kinetic energy is invariant. We first point out that the second order solution, $\omega_2$ itself grows without a bound. Therefore in the second transient stage the decay of the longitudinal velocity $u$ is due to the conservation of the kinetic energy of the sleigh.

\section{Chaotic Dynamics of the sleigh on the attractor}\label{sec:chaos}
\begin{figure}[!h]

 \begin{minipage}{0.48\hsize}
 	\begin{center}
 	 \includegraphics[width =\hsize]{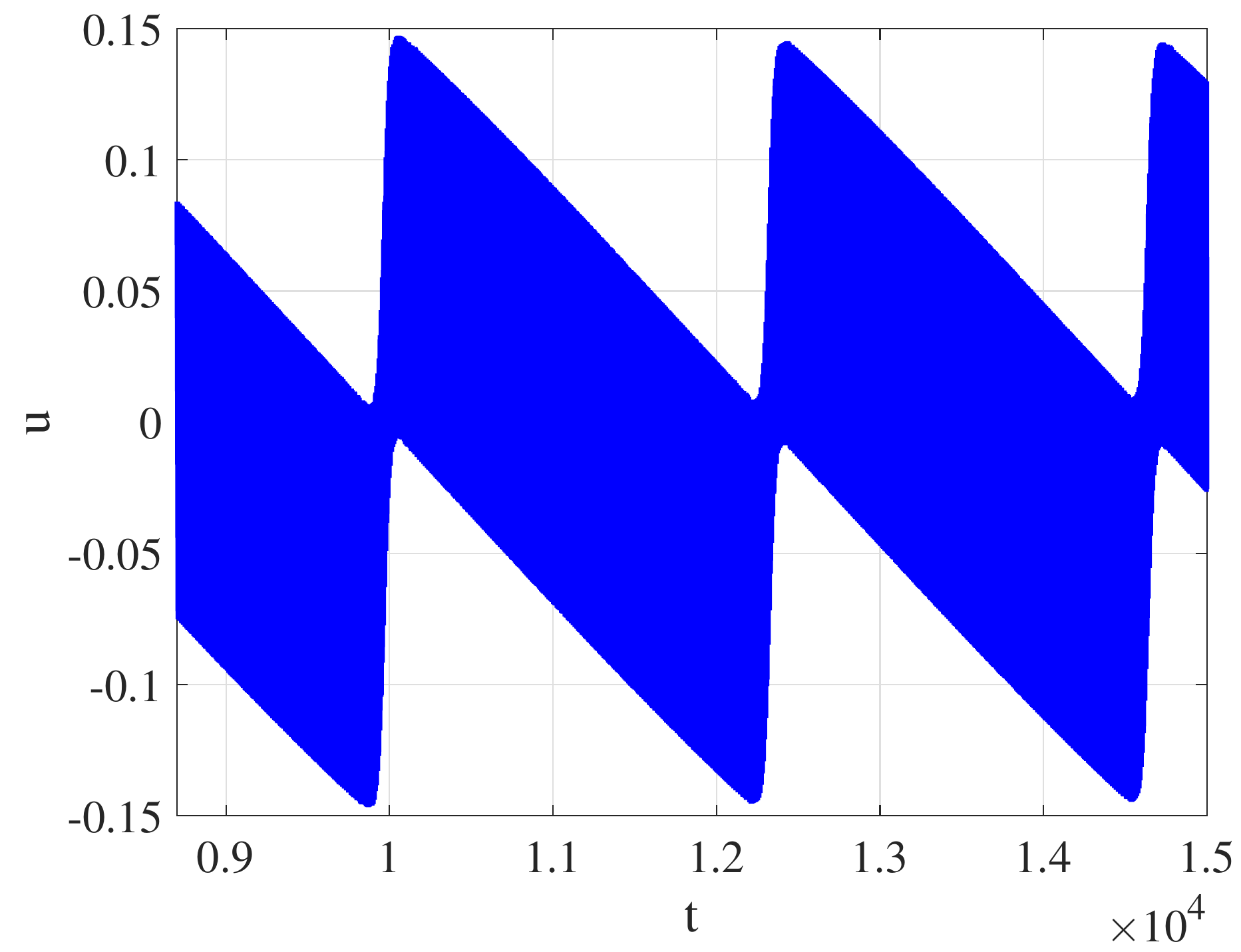}\\
 	   (a)
 	\end{center}
 \end{minipage}
 	 \begin{minipage}{0.48\hsize}
 	 	\begin{center}
 	 		\includegraphics[width =\hsize]{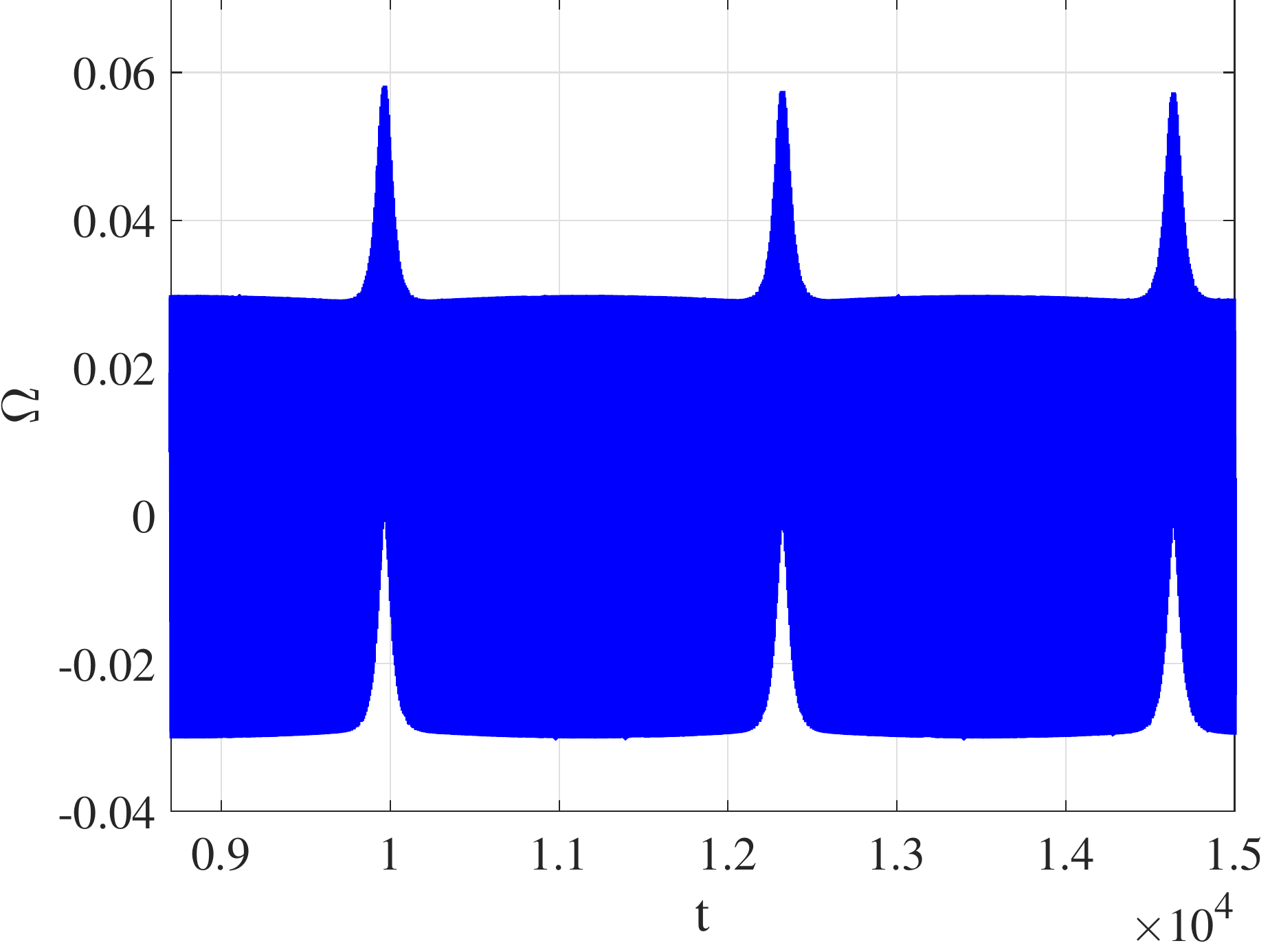}\\
 	 		(b)
 	 	\end{center}
 	 \end{minipage}
 	 
\begin{minipage}{\hsize}
 	\begin{center}
 	 \includegraphics[width =\hsize]{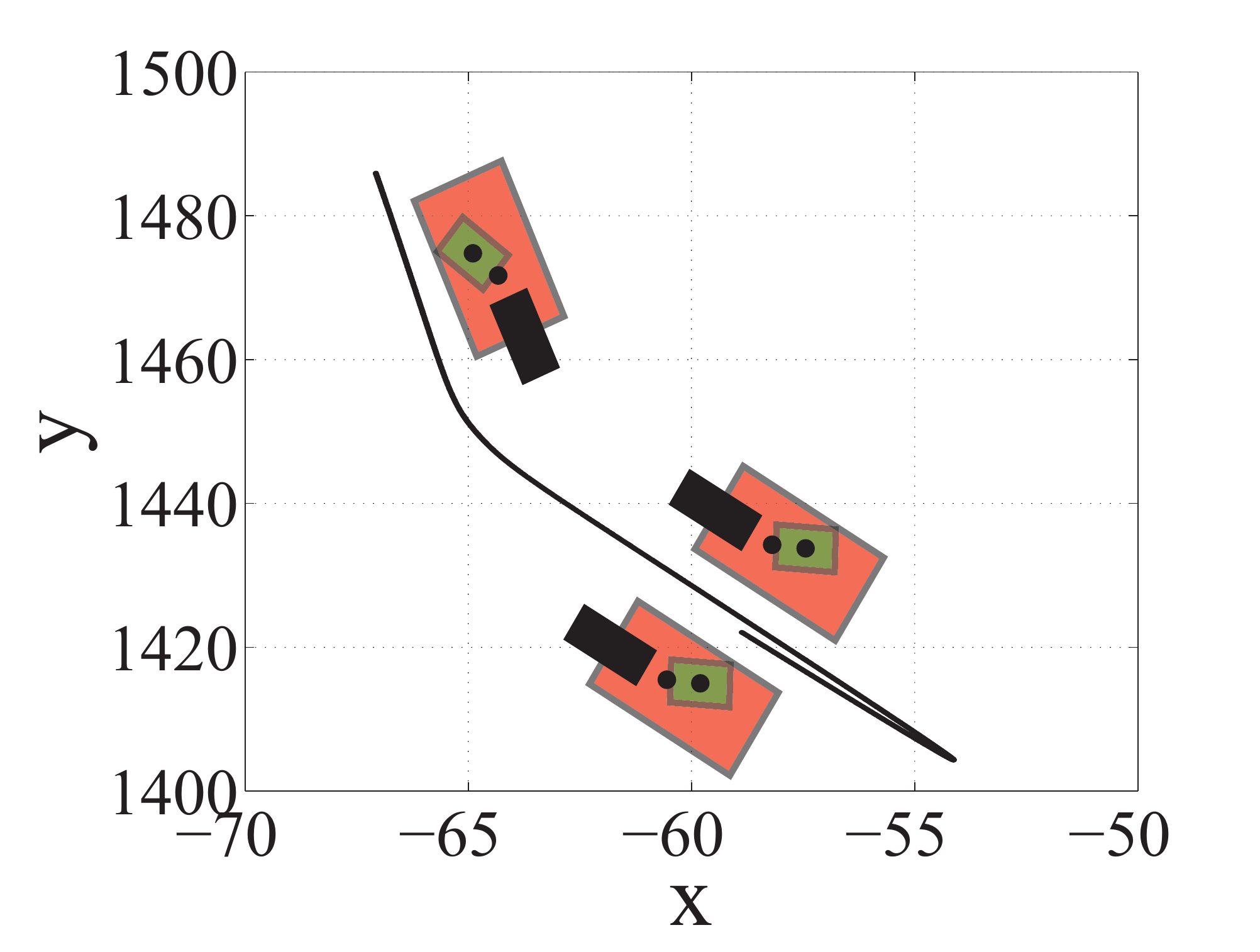}\\
 	   (c)
 	\end{center}
 \end{minipage}

  \begin{minipage}{0.48\hsize}
  \begin{center}
  \includegraphics[width =\hsize]{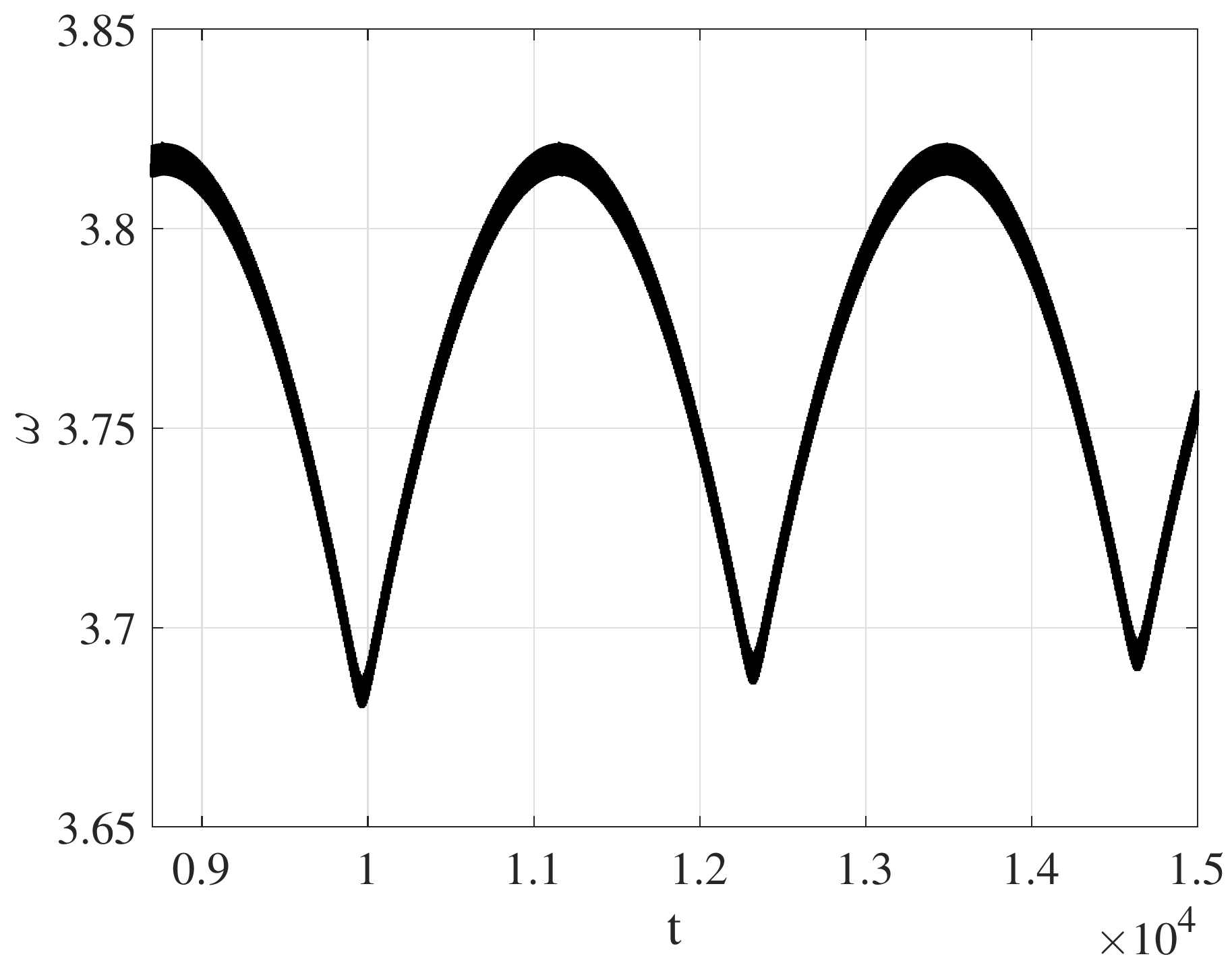}\\
  (d)
  \end{center}
  \end{minipage}
   	 \begin{minipage}{0.48\hsize}
 	 	\begin{center}
 	 		\includegraphics[width =\hsize]{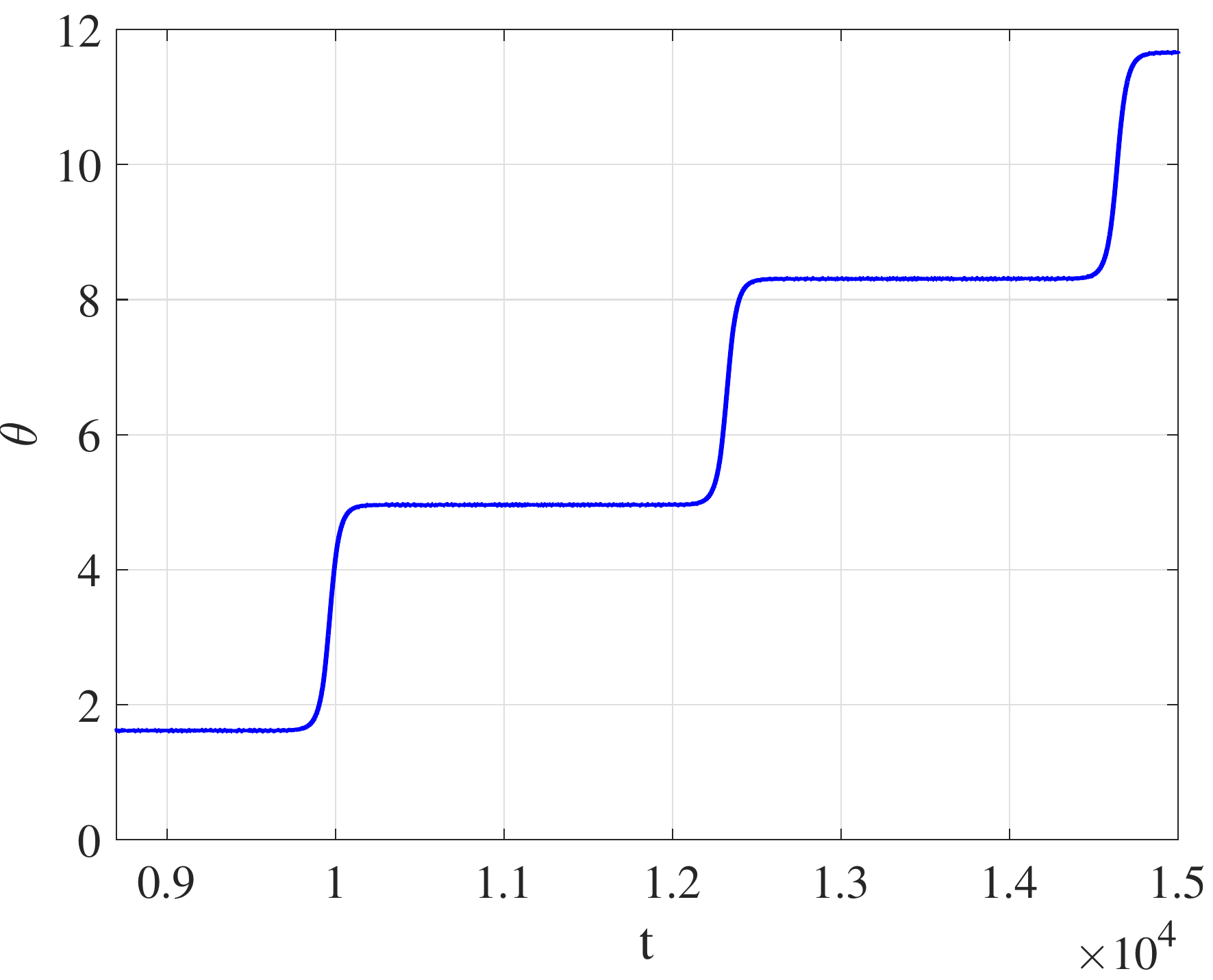}\\
 	 		(e)
 	 	\end{center}
 	 \end{minipage}
    \caption{Steady State behavior of the sleigh. Initial conditions are $u_0=0$, $\Omega=.1$, and $\omega=1$. Subfigures a), b), d) and e) show the parameters of the sleigh over time. Subfigure c) shows the trajectory of the sleigh for the time window $t=9,000s$ to $t=11,500$.}\label{fig:ss}
 \end{figure}

Numerical simulations show that for all initial conditions, except those of a zero measure, the longitudinal velocity of the sleigh converges to a periodic function with multiple frequencies. This is shown in fig. \ref{fig:two_link_long}(a) where from about $t = 8,700$ the velocity $u$ undergoes rapid oscillations, with the mean value of $u$ itself oscillating at a much lower frequency. The angular velocity of the sleigh also undergoes high frequency oscillations with zero mean, along with a spike that occurs between much longer time intervals,  fig. \ref{fig:two_link_long}(b). The angular velocity of the rotor also oscillates with a high frequency with the mean value oscillating at a lower frequency, fig. \ref{fig:two_link_long}(c). We will denote this state of motion of the sleigh as the steady state. To explain this steady state behavior we show in fig. \ref{fig:ss} the kinematic variables over a smaller time window along with the trajectory of the sleigh in the plane during this time window.

In fig. \ref{fig:ss}(a) at about $t = 9,800$, despite the oscillations, the maximum longitudinal velocity of the sleigh is negative, i.e. the cart moves backwards.  As shown in the phase portrait for the zeroth order dynamics, the motion of the Chaplygin sleigh in the backward direction is unstable. Therefore at about $t = 10,000$ $u$ increases, resulting in an increase in magnitude of $\Omega$. This sudden increase in the angular velocity of the sleigh is reflected in the spike in fig. \ref{fig:ss}(b). Figure \ref{fig:ss}(c) shows the trajectory of the sleigh in the physical plane during this spike in angular velocity. In the first transition, between position A and position B, shown in fig. \ref{fig:ss}(c) the sleigh's orientation changes by a large amount (nearly $\pi$ radians) due to the spike in the angular velocity of the sleigh. The motion of the sleigh changes from the backward (in body frame) direction to forward direction, i.e. $u$ becomes positive. At about $t= 12,000$ the longitudinal velocity of the sleigh is once again wholly negative. The average motion of the cart is once again in the backward direction (in the body frame).  What appears to be a sharp turn in the trajectory of the sleigh in the $x-y$ plane is actually the second transition which is characterized by the average longitudinal velocity becoming negative. So the sleigh does not execute a turn at this point, it simply changes the direction of travel.
\begin{figure}[!h]
\begin{center}
\includegraphics[width =\hsize]{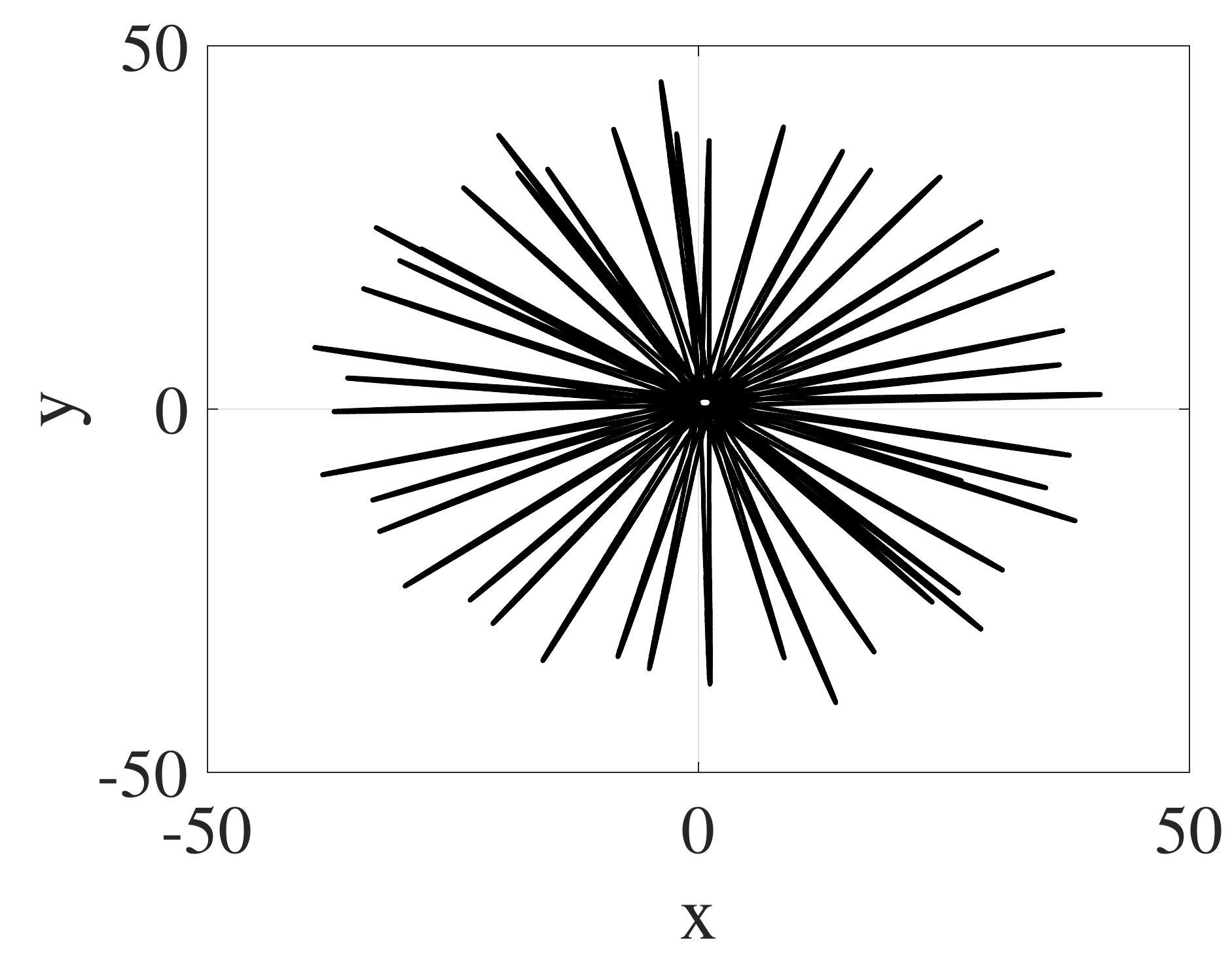}
\end{center}
\caption{Path of the sleigh after $t=100,000s$. Initial conditions are taken after the transient phase to capture the steady state trajectory.}\label{fig:xy_long}
\end{figure}

\begin{figure}[!h]
 \begin{minipage}{0.49\hsize}
 	\begin{center}
 	 \includegraphics[width =\hsize]{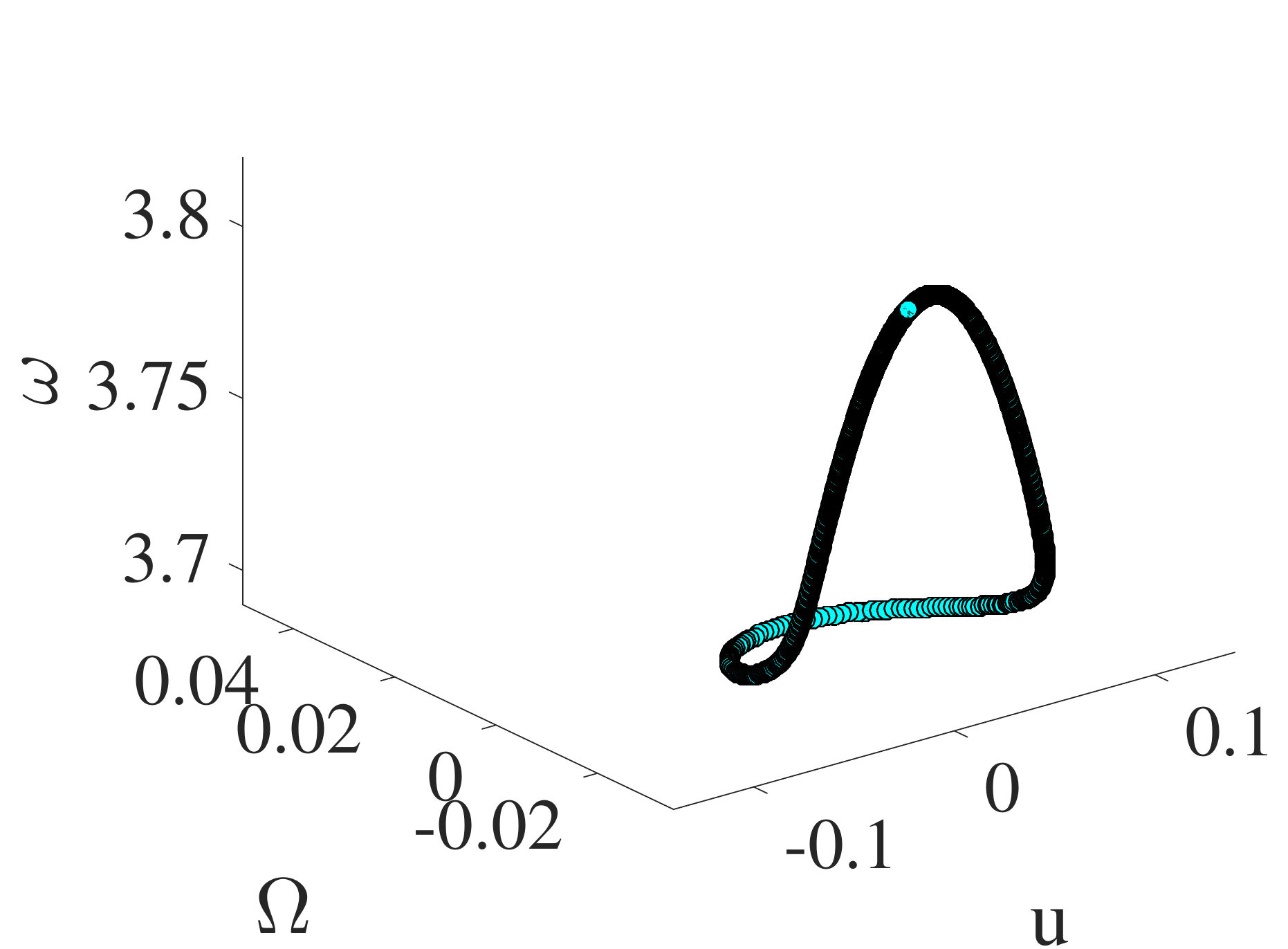}\\
 	   (a)
 	\end{center}
 \end{minipage}
 	 \begin{minipage}{0.49\hsize}
 	 	\begin{center}
 	 		\includegraphics[width =\hsize]{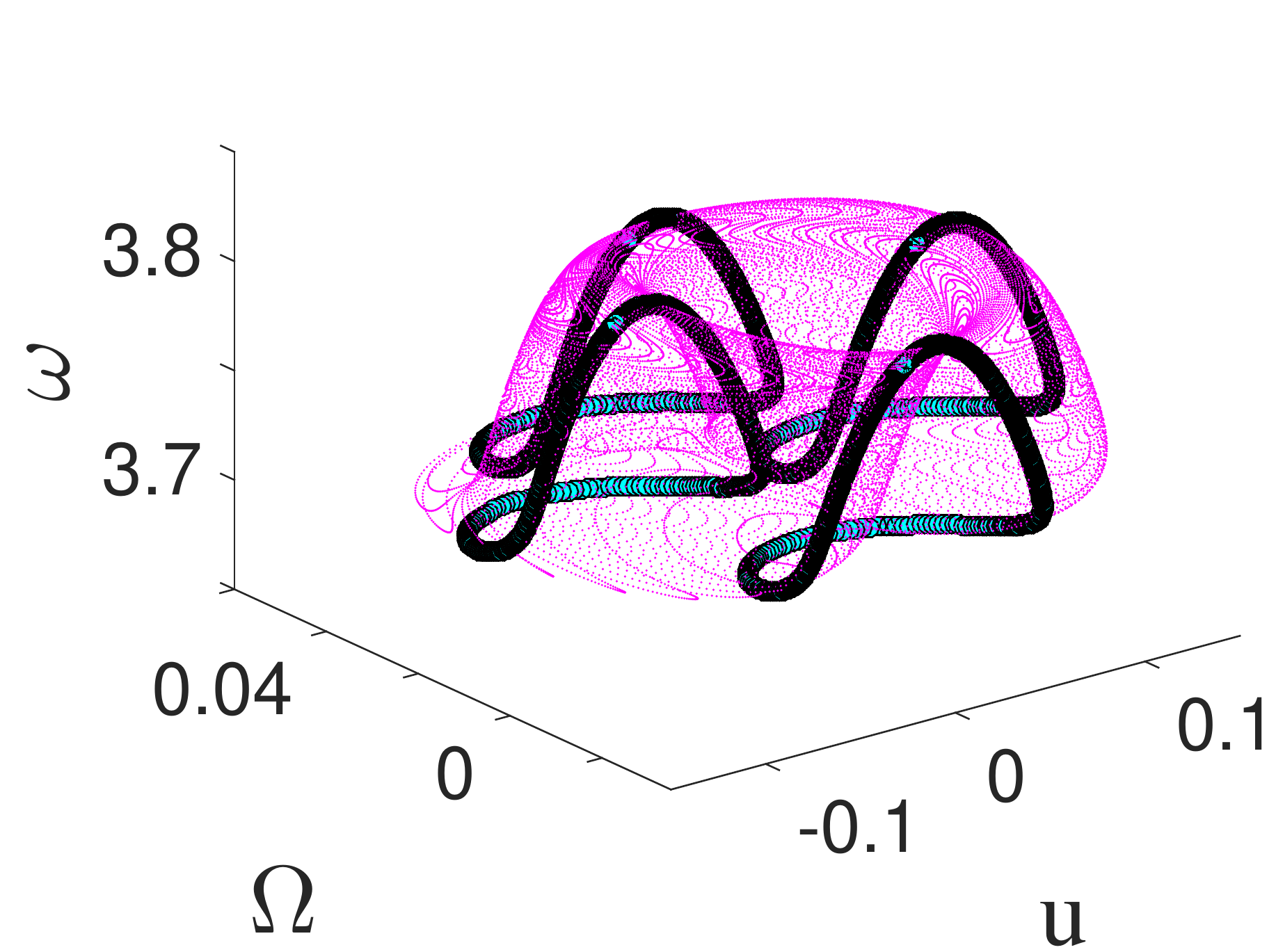}\\
 	 		(b)
 	 	\end{center}
 	 \end{minipage}
 	  \begin{minipage}{0.49\hsize}
 	\begin{center}
 	 \includegraphics[width =\hsize]{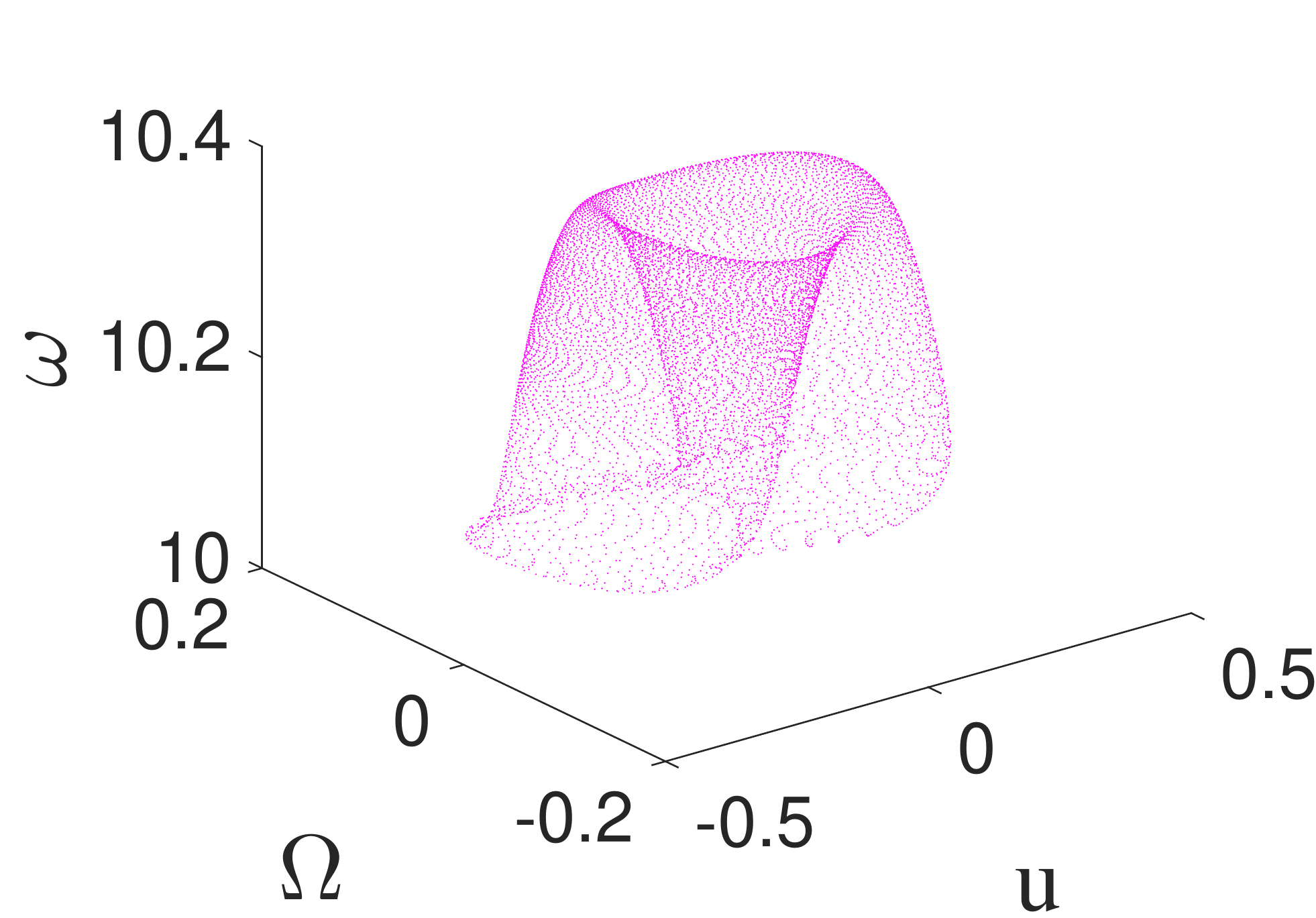}\\
 	   (c)
 	\end{center}
 \end{minipage}
 	 \begin{minipage}{0.49\hsize}
 	 	\begin{center}
 	 		\includegraphics[width =\hsize]{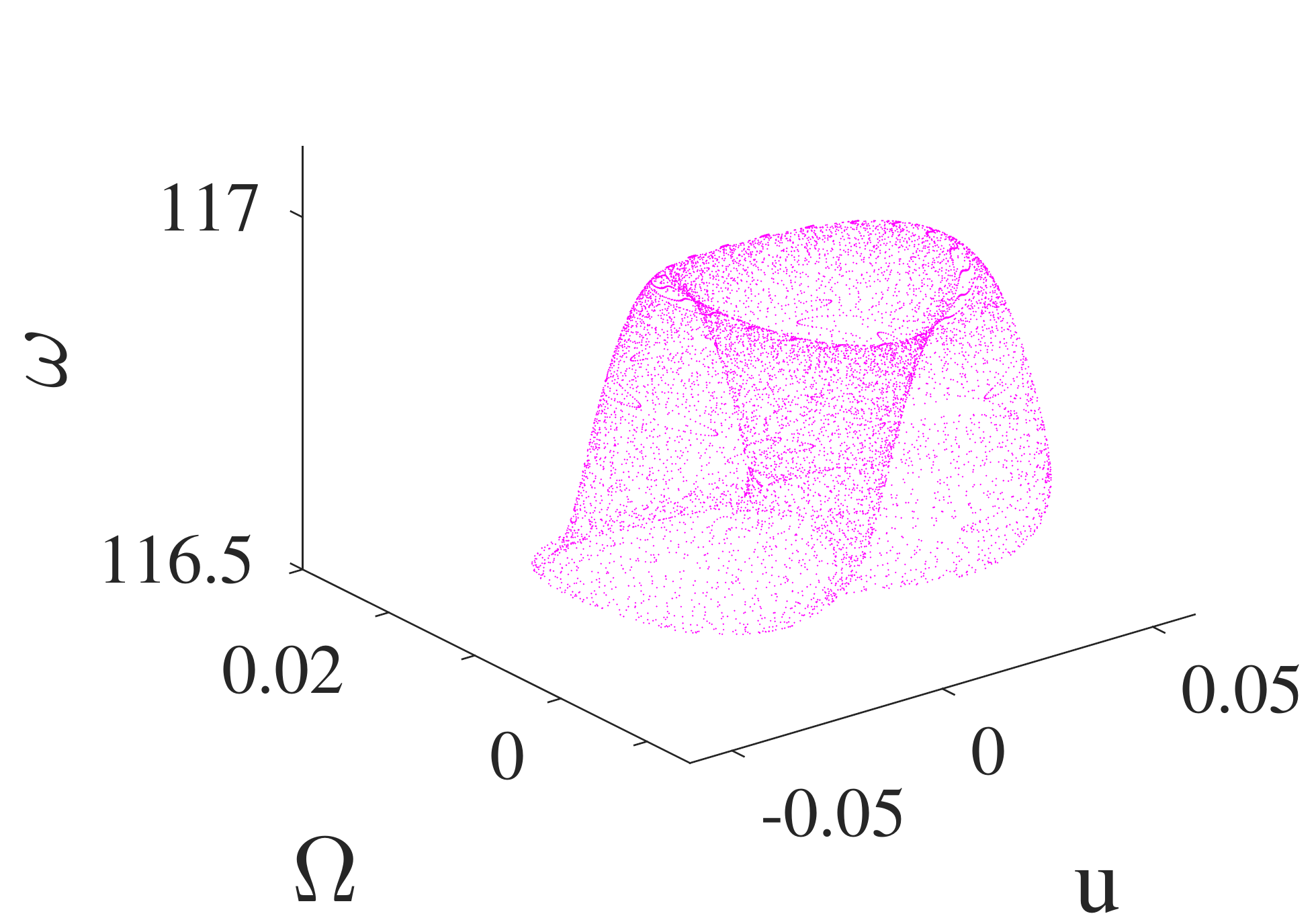}\\
 	 		(d)
 	 	\end{center}
 	 \end{minipage}
 	    \caption{(a) Iterations of the Poincare map $P_{2\pi}$. (b) Iterations of the Poincare map $P_{\pi/2}$ shown by the large filled circles forming four closed loops lie on the attractor $\pi_{\Phi}\mathcal{A}$ shown by the smaller dots, energy is $E_0=0.1127$ and $\epsilon=0.1$ (c) attractor $\pi_{\Phi}\mathcal{A}$ for $\epsilon=0.1$ and an energy of  $E=0.8313$ and (d) for $\epsilon=0.01$ and $E=0.1127$.}\label{fig:attractor}
 \end{figure}

As the mean value of the longitudinal velocity itself undergoes low frequency oscillations (fig. \ref{fig:ss}(a)) the trajectory of the sleigh in the $x-y$ plane undergoes what seem like many sharp turns. Figure \ref{fig:xy_long} shows a generic trajectory of the sleigh in the $x-y$ plane. This trajectory of the sleigh in the $x-y$ plane is in sharp contrast to the trajectory of the Chaplygin sleigh without the internal rotor, whose angular velocity converges to zero for almost all initial conditions. The trajectory of a Chaplygin sleigh  without the internal rotor converges asymptotically to a straight line in the $x-y$ plane and the trajectory is not bounded, see fig. \ref{fig:phase_plot}(b). In contrast the trajectory of the sleigh with even a small passive rotor is bounded as shown in fig. \ref{fig:xy_long}. Moreover numerical simulations suggest that the path of the sleigh is not periodic. To illustrate the non periodic nature of the trajectories of the dynamical system  \eqref{eq:u}-\eqref{eq:delta_dot}, we consider the Poincare section, $\Sigma_{2\pi} =\{u,\Omega,\omega,\delta = 2n\pi\}$ where $n=0,1,2,3...$ and the first return Poincare map to the Poincare section, 
\begin{equation}
P_{2\pi}: (u_n, \Omega_n, \omega_n, 2n\pi) \mapsto \Phi_{t_n}^{t_{n+1}}(u_n, \Omega_n, \omega_n, 2(n+1)\pi).
\end{equation}
Figure \ref{fig:attractor}(a) shows the Poincare map for a large number of iterations. The Poincare map is not periodic for thousands of iterations and fills out a closed curve. We further consider the Poincare section, $\Sigma_{\pi/2} =\{u,\Omega,\omega,\delta = n\pi/2\}$ and the first return Poincare map, $P_{\pi/2}$ to this section $\Sigma_{\pi/2}$. figure \ref{fig:attractor}(b) shows a large number of iterations of this map, which form four distinct closed loops. In the steady state the trajectory of the dynamical system \eqref{eq:u}-\eqref{eq:delta_dot} lies on an attractor $\mathcal{A}$. The iterations of these Poincare maps lie on the projection of the attractor, $\pi_\mathcal{A} \subset \mathbb{R}^3 $ which is the projection of the flow map  $\pi_{\Phi} : \mathbb{R}^3\times S^1 \mapsto \mathbb{R}^3$. The attractor, $\pi_\mathcal{A}$, is shown by the small dots in fig. \ref{fig:attractor}(b).

The reduced equations of motion of the sleigh, \eqref{eq:u} - \eqref{eq:delta_dot} are dissipative in the sense of decreasing phase space volumes. The trace of the Jacobian obtained by linearizing these reduced velocity equations is nonzero for almost any $(u,\Omega, \omega, \delta)$.  However the kinetic energy of the sleigh is an invariant. The manifold $\mathcal{M}$ is foliated by level sets of the kinetic energy. Trajectories that lie on a level set of the kinetic energy converge to an attractor that is a subset of this level set. The attractor and its projection shown in fig.\ref{fig:attractor}(b) are subsets of an invariant level set of the kinetic energy. The attractor is dependent on the kinetic energy of the sleigh as well as the value of $\epsilon$. However the topology of the attractor persists across a broad range of values of $\epsilon$, and $E$. For example fig. \ref{fig:attractor}(c) shows the attractor for  a different energy $E = 0.8313$ and the same $\epsilon =0.1$ while fig. \ref{fig:attractor}(d) shows the attractor for the same energy and $\epsilon = 0.01$. For the combination of larger values of $\epsilon$ and smaller values of total energy $E$, the attractor $\pi_{\mathcal{A}}$ can be flatter with smaller variations in the range of the $\omega$.

The dynamics on the attractor have sensitive dependence on initial conditions. This can be seen through the computation of the Lyapunov exponents. Adopting the algorithm proposed in \cite{swinney_1985}, we computed the four Lyapunov exponents for the dynamical system, \eqref{eq:u}-\eqref{eq:delta}. The four Lyapunov exponents are $2.9e^{-4}$, $2.203e^{-4}$, $-2.2e^{-4}$ and $2.02e^{-5}$. The Lyapunov exponents were computed for a simulation time of $4000$s at intervals of $0.01$s. A pre simulation for an initial run time of $10000$s performed first to allow the transient dynamics to decay to very small values. We verified the convergence of the computation of the Lyapunov exponents by checking the variation in the computed values of these exponents. This variation of the Lyapunov exponents between time steps oscillated between $7\times10^{-8}$ and $-9\times 10^{-8}$ for the last $200$s of the simulation. This variation is of the order of $0.01\%$ in the values of the positive Lyapunov exponents. 

\begin{figure}

\begin{center}
\includegraphics[width =\hsize]{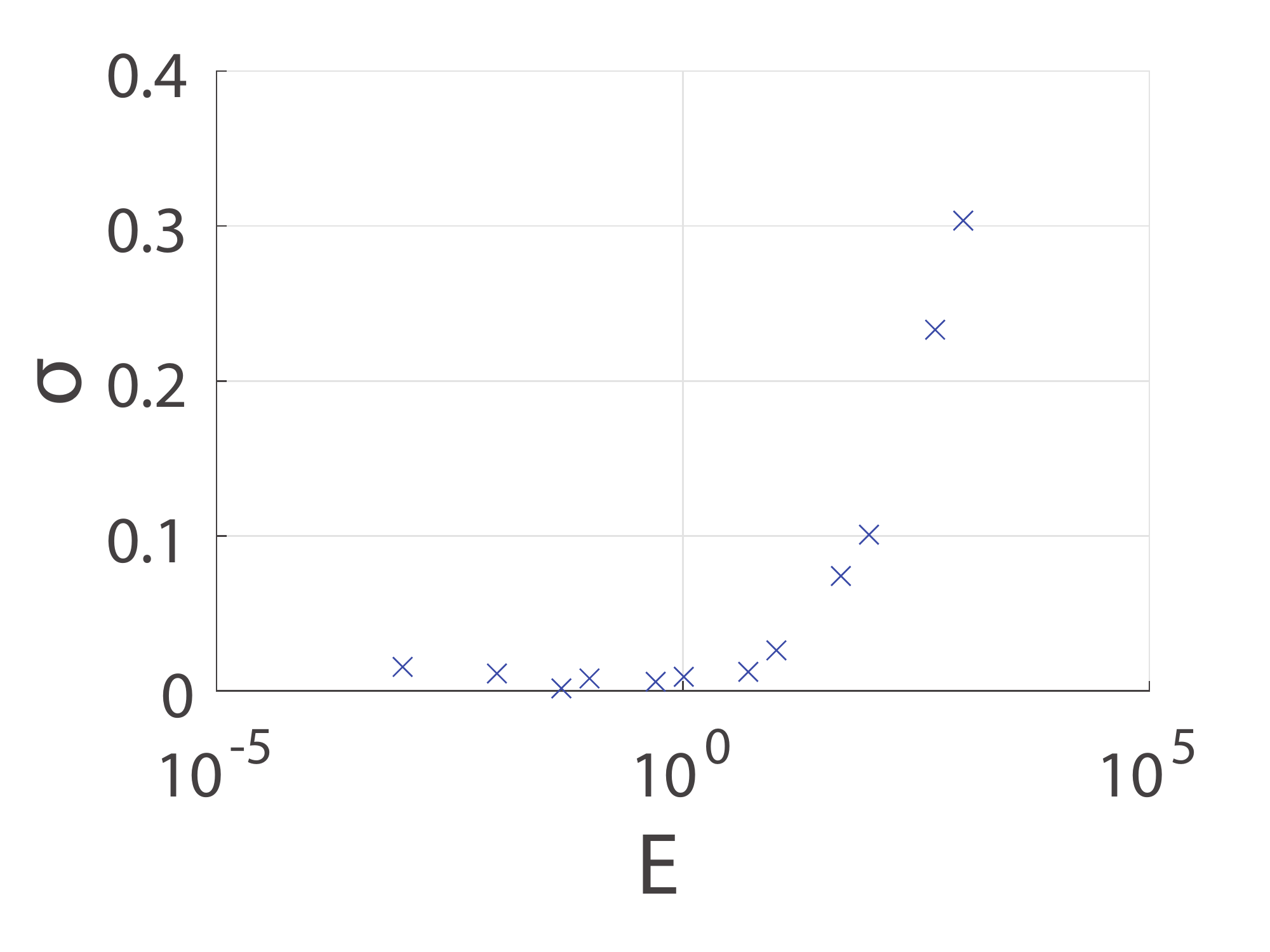}
\end{center}

\caption{Maximum Lyapunov exponent for energies ranging from $E=10^{-3}$ to $E=10^{3}$ with $\epsilon=0.1$. The smallest value of leading Lyapunov exponent in the above graph is about  $1.1\times10^{-4}$.}\label{fig:lyap}
\end{figure}
\begin{figure}

\begin{center}
\includegraphics[width =\hsize]{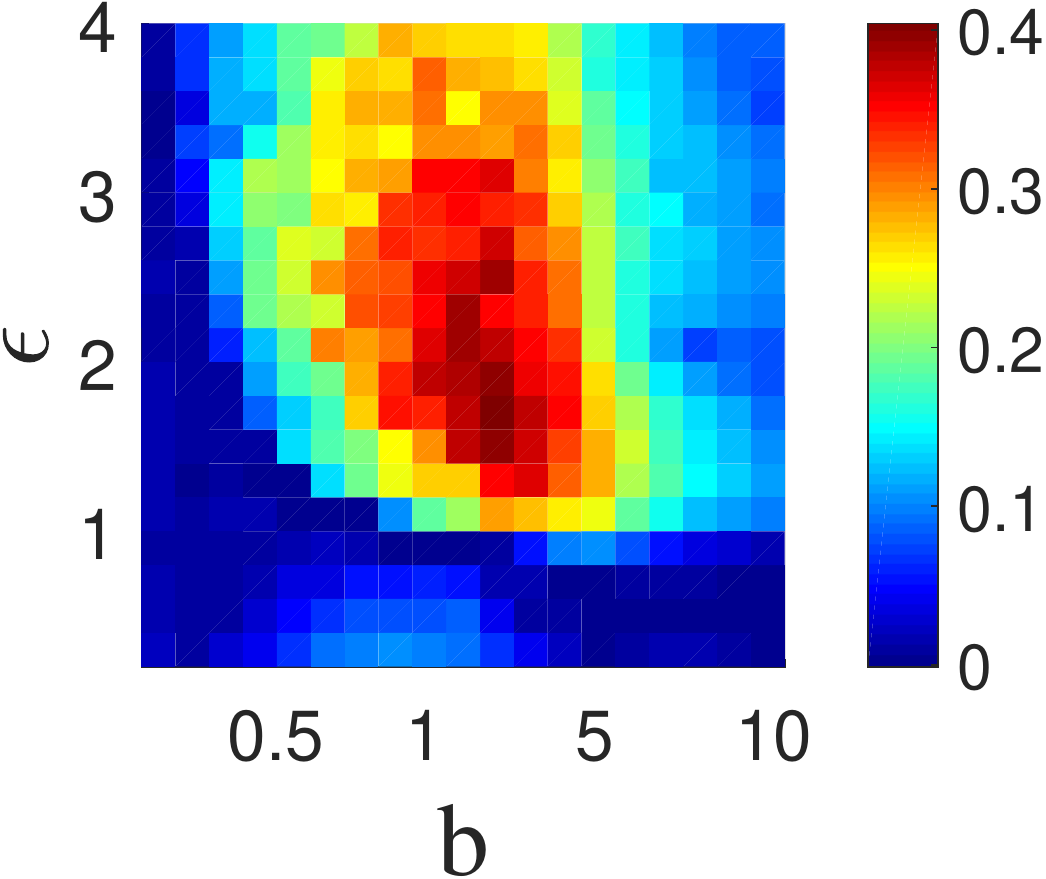}
\end{center}

\caption{Maximum Lyapunov exponent for $E=0.1127$ for a range of $\epsilon$ and $b$. The largest Lyapunov exponent is always positive.}\label{fig:LE}
\end{figure}

The leading Lyapunov exponents are calculated for different energy levels, $E$, holding the values of $\epsilon = 0.1$ and $b = 1$, to verify that the chaotic behavior does not depend on the energy on the system. The highest Lyapunov exponent plotted against the energy is shown in fig. \ref{fig:lyap}. The highest Lyapunov exponent increases with the energy associated with the system and the LE is positive for energy as low as $10^{-4}$.   Similarly changes to the parameters $\epsilon$ and $b$ across a range of values of the energy $E$ produce chaotic dynamics, with the largest Lyapunov exponent always being positive. The leading Lyapunov exponent is shown in fig. \ref{fig:LE} for a range of values of $b$ and $\epsilon$ with $E=0.1127$. The smallest value of the leading Lyapunov exponent in fig. \ref{fig:LE} is greater than $10^{-4}$.
The calculations of the Lyapunov exponents strongly suggest the existence of sensitive dependence of initial conditions, positive Lyapunov exponents and a chaotic attractor for the dynamical system \eqref{eq:u}-\eqref{eq:delta_dot} for a broad range of parameters $\epsilon$, $b$ and $E$. 

\begin{figure}[!h]
 \begin{minipage}{0.49\hsize}
 	\begin{center}
 	 \includegraphics[width =\hsize]{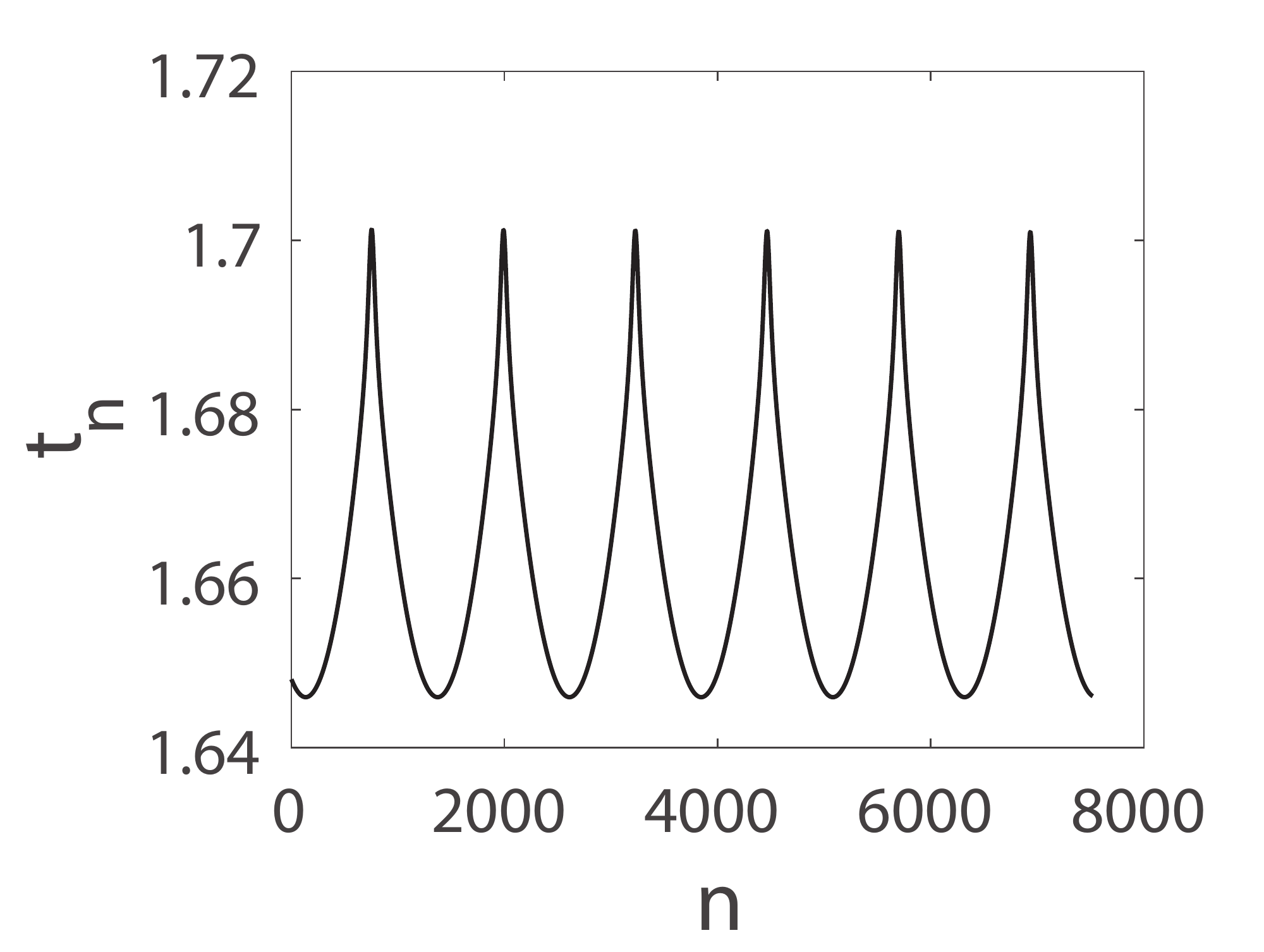}\\
 	   (a)
 	\end{center}
 \end{minipage}
 	 \begin{minipage}{0.49\hsize}
 	 	\begin{center}
 	 		\includegraphics[width =\hsize]{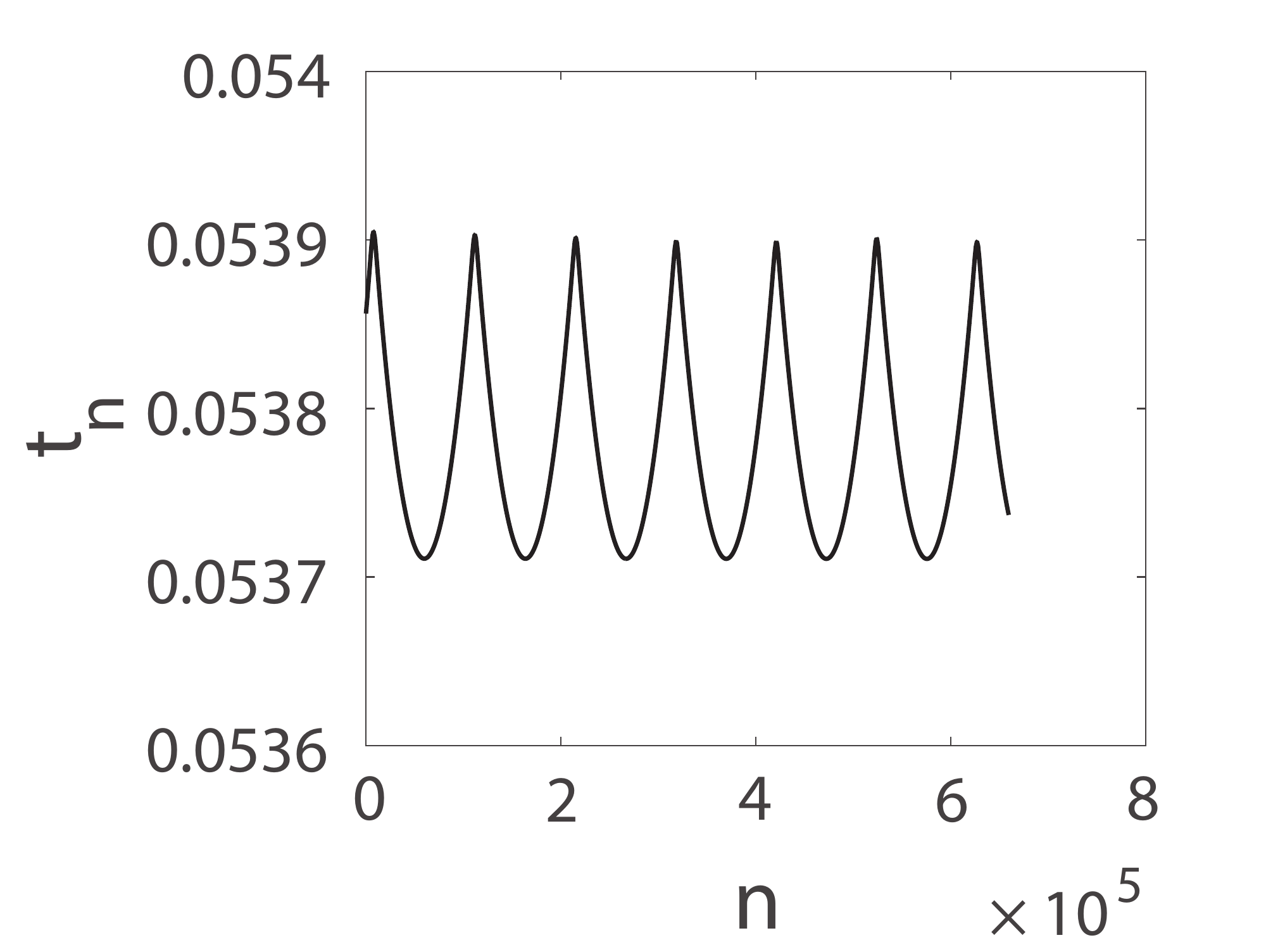}\\
 	 		(b)
 	 	\end{center}
 	 \end{minipage}
 	    \caption{Return times $t_n$ of the Poincare map $P_{2\pi}$ for (a) $\epsilon=0.1$ and (b) $\epsilon=0.01$ with an energy of $E=0.1127$. }\label{fig:tc}
 \end{figure}

In order to further verify the aperiodic behavior of flow of \eqref{eq:u}-\eqref{eq:delta_dot} we plot the return times for the map $P_{2\pi}$. The return times for two different values of $\epsilon$ are shown in fig. \ref{fig:tc} for a large number of iterations of the map $P_{2\pi}$ are shown. The return times for both the cases in fig. \ref{fig:tc}  are such that they are bounded in an interval. Furthermore, no two return times are the same upto a precision of $10^{-7}$. Return time  computations for the map $P_{2\pi}$ for a broad range of values of $\epsilon$, $E$ and $b$ show a similar behavior, suggesting the presence of a large or infinite number of frequencies for the function $\delta(t)$.

\begin{figure}[!h]
\begin{minipage}{\hsize}
 	\begin{center}
 	 \includegraphics[width =\hsize]{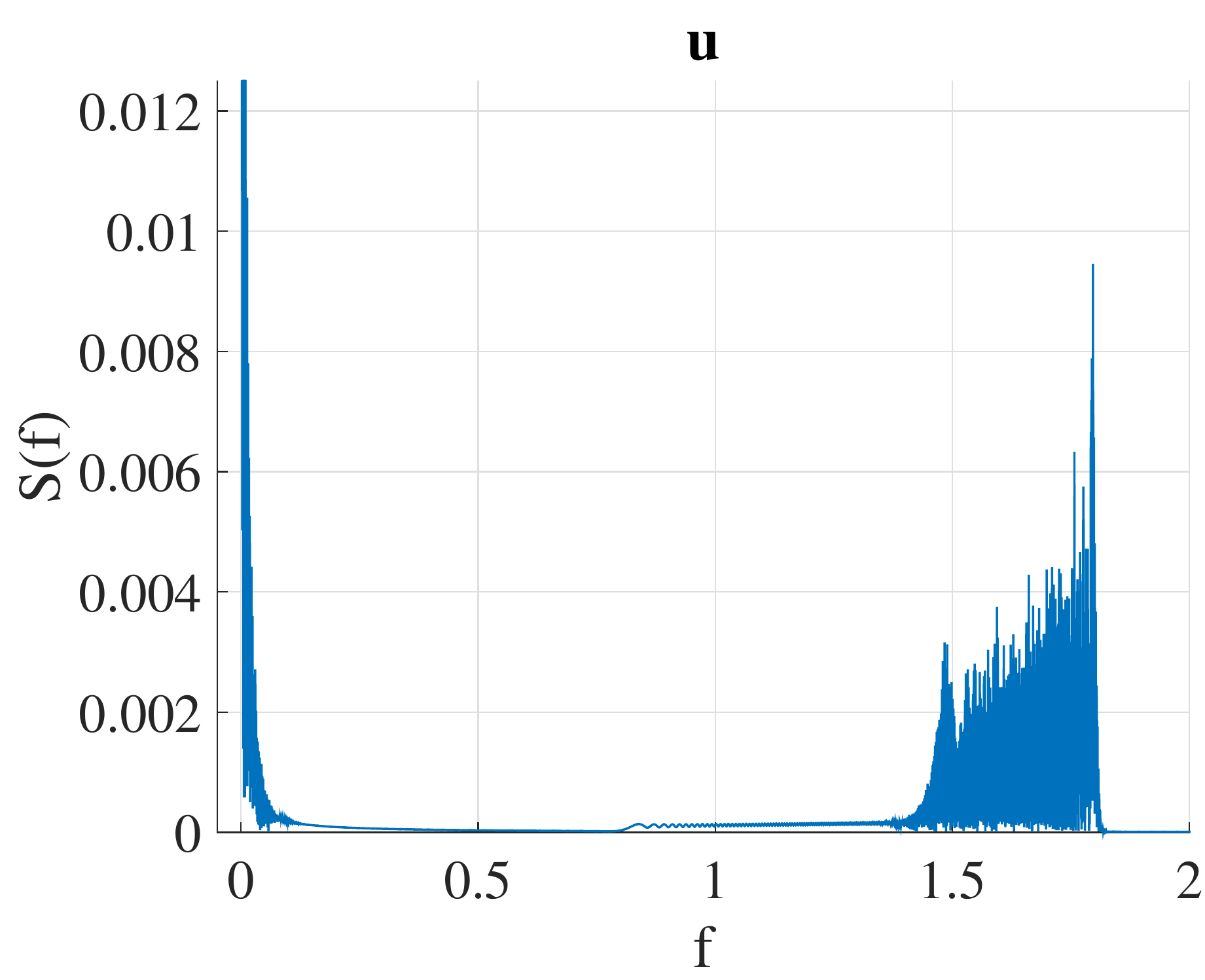}\\
 	\end{center}
 \end{minipage}
 
 	 \begin{minipage}{\hsize}
 	 	\begin{center}
 	 		\includegraphics[width =\hsize]{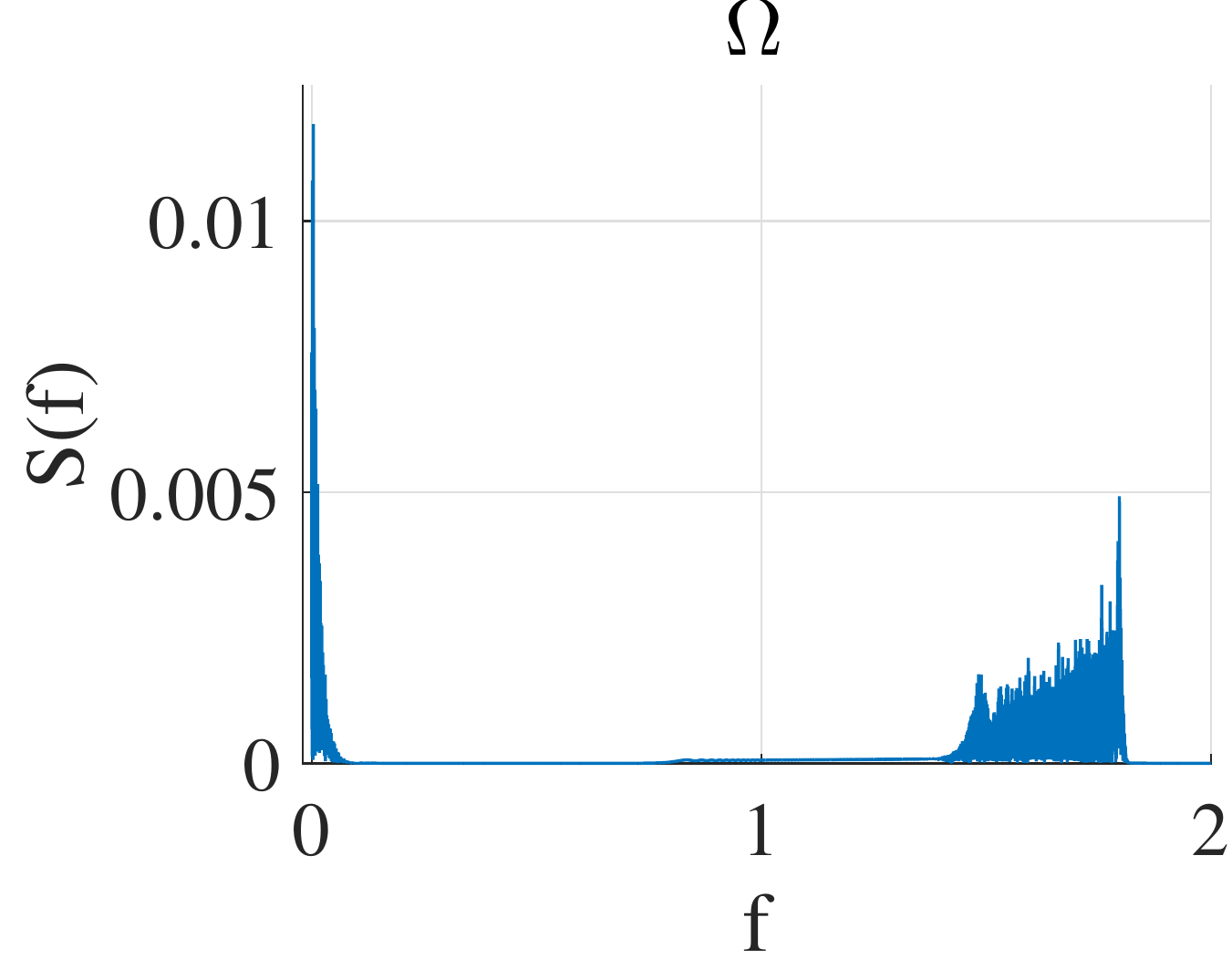}\\
 	 	\end{center}
 	 \end{minipage}
   \caption{Power spectral density plots for $u$ and $\Omega$. Amplitude $S(f)$ is plotted against frequency $f$.  $\epsilon=0.1$ and $E=0.1127$.\label{fig:psd1}}
 \end{figure}

\begin{figure}[!h]
\begin{minipage}{\hsize}
 	\begin{center}
 	 \includegraphics[width =\hsize]{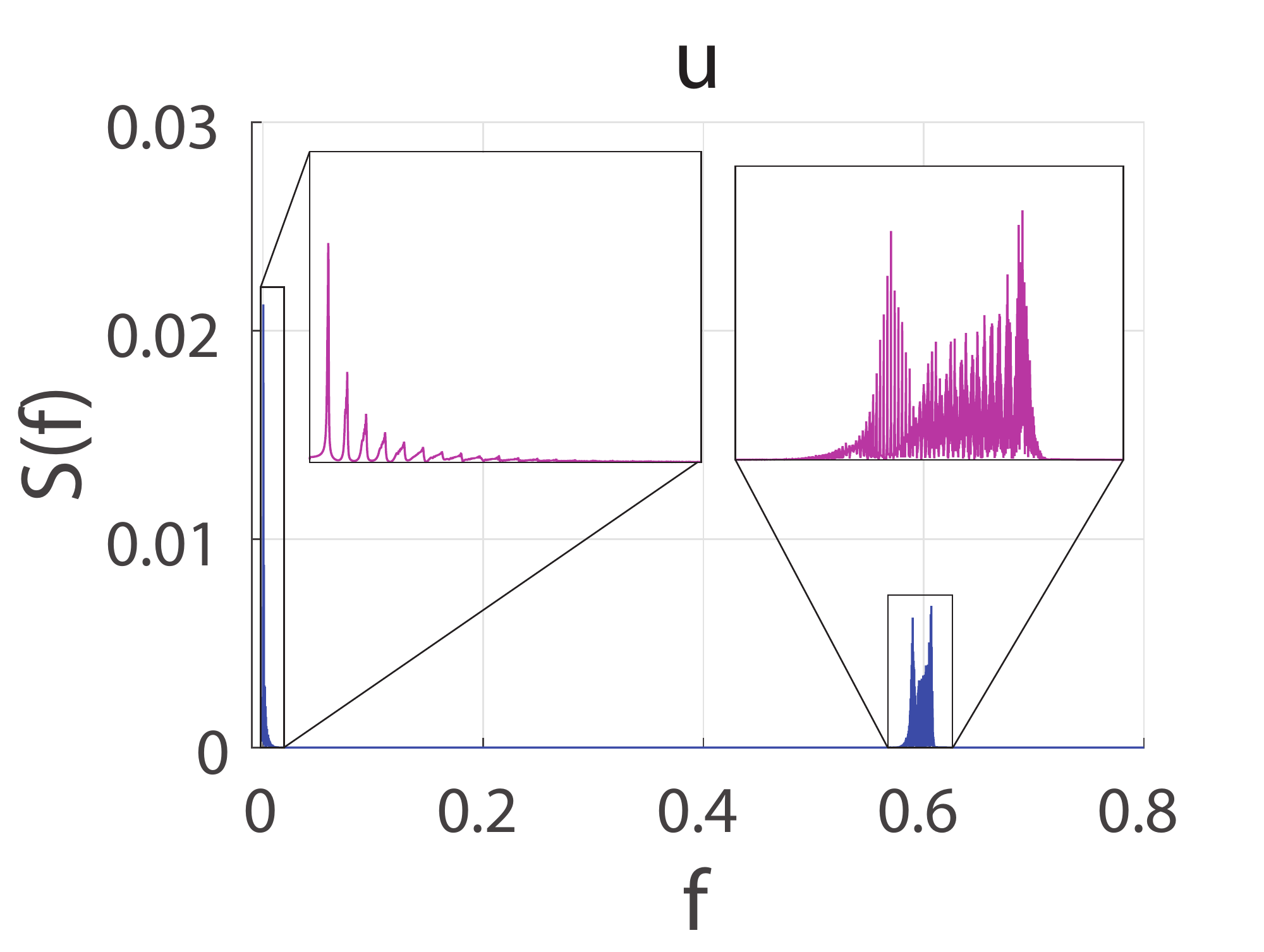}\\
 	\end{center}
 \end{minipage}

 	 \begin{minipage}{\hsize}
 	 	\begin{center}
 	 		\includegraphics[width =\hsize]{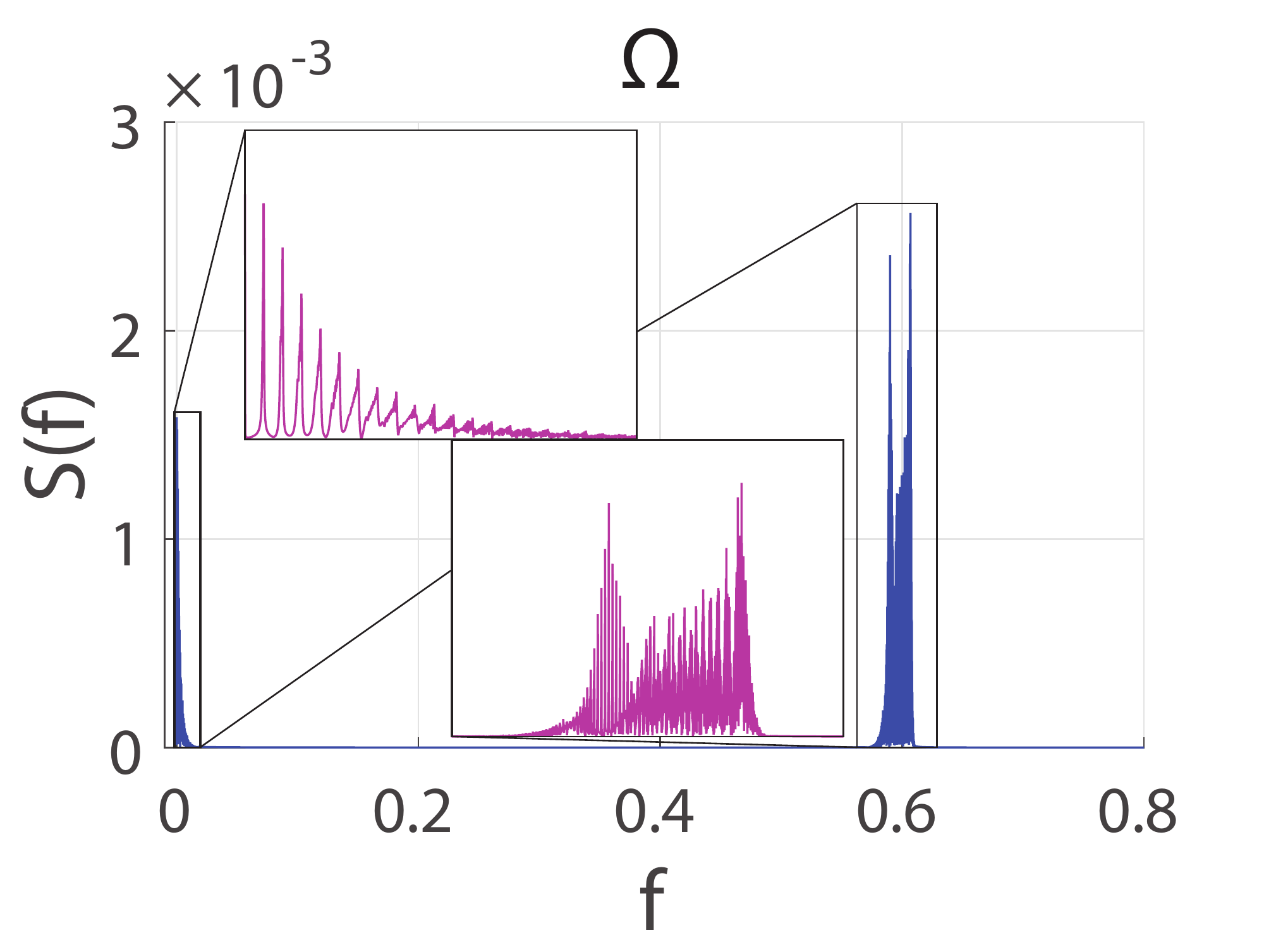}\\
 	 	\end{center}
 	 \end{minipage}
   \caption{Power spectral density plots for $u$ and $\Omega$. Amplitude $S(f)$ is plotted against frequency $f$.  $\epsilon=0.01$ and $E=0.1127$.\label{fig:psd}}
 \end{figure}

 The dynamics on the attractor are also aperiodic. A power spectral density plot of $u(t)$ and $\Omega(t)$ reveal that a very large number of frequencies are present clustered into two regions of the frequency spectrum. These plots for the case of $\epsilon = 0.1$ is shown in fig. \ref{fig:psd1} and for the case of $\epsilon = 0.01$ in fig. \ref{fig:psd}. In fig. \ref{fig:psd1}(a)-(b) the clustering of the power spectrum for the $u(t)$ and $\Omega(t)$ around the frequency zero has many distinct well defined peaks, which broaden as the frequency increases, suggestive of quasiperiodic behavior. The power spectrum away from zero is clustered in a broadband on a frequency interval that is approximately $f\in [1.45, 1.8]$. The frequency spectra are computed in MATLAB using the $fft$ function with an input signal on a time interval $5000$s with time steps of $10^-3$. As in the computation of the Lyapunov exponents a pre simulation for $10000$s was performed to allow the transient dynamics to decay to negligible values.
 
 For a smaller value of $\epsilon = 0.01$, fig. \ref{fig:psd}, the frequency spectra of $u(t)$ and $\Omega(t)$ have a narrower  the frequency interval $[0.56, 0.6]$ in which a very large number of frequencies are present in a continuous band. At the zero frequency the frequency spectra has many but distinct peaks, suggestive of quasiperiodic behavior in this small frequency band.

The power spectrum plots for $u(t)$ and $\Omega(t)$ together with the return times suggest that within a small interval all frequencies are present demonstrating the aperiodic behavior. At the same time, the presence of fat peaks in the frequency spectra for smaller $\epsilon$ (fig. \ref{fig:psd}) along with the the smaller difference in the return times of the map $\P_{2\pi}$ (fig. \ref{fig:tc}) suggest a possible quasiperiodic route to chaotic behavior. Such chaotic behavior can be verified numerically for values of  $\epsilon$ as small as $10^{-5}$. However the numerics become unreliable for as the parameetr $\epsilon$ becomes smaller. Similarly for very large values of $\epsilon$, exceeding 20, the leading Lyapunov exponents become smaller than $10^{-5}$ suggesting that the chaotic behavior could disappear in atleast some windows as the parameters of the system vary.

An analytical approach to the possible bifurcations in the dynamical system \eqref{eq:u}-\eqref{eq:delta_dot} present many challenges because of the singular nature of $\epsilon$. In the absence of the internal rotor, i.e. $\epsilon = 0$, the dynamical system \eqref{eq:u}-\eqref{eq:delta_dot} reduces to a two dimensional system.  The solution to this case is that the angular velocity of the sleigh, $\Omega$, decays to zero and the longitudinal velocity of the sleigh, $u$, converges to a constant value. For any small but finite $\epsilon$, the dimension of the system increases to four. This singular nature of $\epsilon$ does not allow a traditional analysis of bifurcations (if any) of the invariant sets of the dynamical system \eqref{eq:u}-\eqref{eq:delta_dot} and the route to chaos.  The numerical analysis however indicates that the system follows a quasiperiodic route to chaos around $\epsilon = 0$.
 
\section{Conclusion and Discussion}
In this paper we analyzed the dynamics of the Chaplygin sleigh with a passive internal rotor. We found that the sleigh exhibits very rich and nonintuitive dynamics with the addition of an internal degree of freedom. The motion of the classical Chaplygin sleigh without a passive rotor approaches a straight line asymptotically. The angular velocity of the sleigh decays to zero for almost all initial conditions. With the addition of a passive internal rotor, the motion of the sleigh becomes chaotic for almost all initial conditions. The trajectory of the sleigh in the $x-y$ plane is bounded but not periodic. 

The significant effect that a passive internal degree of freedom can have on a nonholonomic system  inspires questions of passive mechanical control of robotic systems. The potential applications are geared towards robots with nonholonomic constraints which are not purely kinematic.  For instance the trajectory of the Chaplygin sleigh shown in fig. \ref{fig:xy_long} can be used as a basis for the problem of coverage, without additional sensing and control. This is very interesting because even a small rotor can change the dynamics of the Chaplygin sleigh from an asymptotic straight line motion in the plane to the possibly chaotic and ergodic motion in the plane. The addition of dissipative and stiff elements can produce limit cycles such as those observed in \cite{ft_nd_2018} which could passively direct the dynamics of the robot to a desired invariant state.  The ability of switching on or off an internal degree of freedom could further offer efficient ways to control the motion of robots with nonholonomic constraints.

\section{Acknowledgements}
This paper is based upon work supported by the National Science Foundation under grant number CMMI 1563315.


\end{document}